\documentclass[aps,pra,reprint,superscriptaddress]{revtex4-1}
\usepackage{graphicx}
\usepackage{bm, amsmath, amssymb}
\usepackage{color}
\usepackage{times}
\newcommand{\beq}{\begin{equation}}
\newcommand{\eeq}{\end{equation}}
  \newcommand{\beql}[1]{\begin{equation}\label{eq:#1}}
 \newcommand{\beqa}{\begin{eqnarray}}
  \newcommand{\eeqa}{\end{eqnarray}}
  \newcommand{\C}{\mathbb{C}}
 \newcommand{\R}{\mathbb{R}}
  \newcommand{\M}{{\bf M}}
 \newcommand{\bM}{\mathbf{M}}
  \newcommand{\bP}{\mathbf{P}}
  \newcommand{\bS}{\mathbf{S}}
  \newcommand{\bx}{\mathbf{x}}
 \newcommand{\cG}{\mathcal{G}}
  \newcommand{\cH}{\mathcal{H}}
  \newcommand{\cK}{\mathcal{K}}
  \newcommand{\al}{\alpha}

  \newcommand{\da}{\dagger}
  
  \newcommand{\ep}{\varepsilon}
  \newcommand{\et}{\eta}

  \newcommand{\la}{\lambda}
  \newcommand{\nn}{\nonumber}

  \newcommand{\ps}{\psi} 
  \newcommand{\si}{\sigma} 
  \newcommand{\ta}{\tau}
  \newcommand{\De}{\Delta}                                          
  \newcommand{\Eq}[1]{Eq.~(\ref{eq:#1})}                                         
  \newcommand{\Tr}{\mbox{\rm Tr}}
  \newcommand{\eq}[1]{(\ref{eq:#1})}
\newcommand{\bra}[1]{\langle#1|}
\newcommand{\ket}[1]{|#1\rangle}
\newcommand{\ketbra}[1]{\ket{#1}\bra{#1}}
\newcommand{\epg}{\ep_{G}}
\newcommand{\epn}{\ep_{{\rm NO}}}
\newcommand{\epu}{\overline{\ep}}
\renewcommand{\t}{t_0}
\newcommand{\av}[1]{\langle #1 \rangle}
\renewcommand{\Re}{{\rm Re\,}}
\renewcommand{\Im}{{\rm Im\,}}
\newcommand{\bmat}{\left(\begin{array}{cc}}
\newcommand{\emat}{\end{array}\right)}
\newcommand{\bvec}{\left(\begin{array}{r}}
\newcommand{\evec}{\end{array}\right)}
\newcommand{\AM}{{}}
\newcommand{\PM}{\Pi}
\newcommand{\hPM}{\hat{\Pi}}
\newcommand{\TM}{T^{*}}
\newcommand{\RS}{\rho}
\newcommand{\RSX}{\RS\otimes\ketbra{\xi}}
 \newcommand{\kxi}{\vert \xi\rangle}
 \newcommand{\bA}{\mathbf{A}}
\def\1{\mathchoice{\rm 1\mskip-4.2mu l}{\rm 1\mskip-4.2mu l}{\rm
        1\mskip-4.6mu l}{\rm 1\mskip-5.2mu l}}
\newcommand{\cI}{\mathcal{I}}
\newcommand{\deq}[1]{\begin{align}#1\end{align}}
\newcommand{\deqs}[1]{\begin{align*}#1\end{align*}}

\newcommand{\red}[1]{\color{red}{#1}\color{black}}
\begin{document}
\title{Error-disturbance relation in Stern-Gerlach measurements}
\author{Yuki Inoue}
\affiliation{Graduate School of Informatics, Nagoya University,
Chikusa-ku, Nagoya, 464-8601, Japan}
\email[]{inoue.y.at@gmail.com}
\author{Masanao Ozawa}
\affiliation{Graduate School of Informatics, Nagoya University,
Chikusa-ku, Nagoya, 464-8601, Japan}
\affiliation{College of Engineering, Chubu University,
1200 Matsumoto-cho, Kasugai-shi, Aichi, 487-8501, Japan}
\email[]{ozawa@is.nagoya-u.ac.jp}

\begin{abstract}
Although Heisenberg's uncertainty principle is represented by a rigorously proven relation 
about intrinsic uncertainties in quantum states, Heisenberg's error-disturbance relation (EDR)
has been commonly believed to be another aspect of the principle.
Based on the recent development of universally valid reformulations of
Heisenberg's EDR, 
we study the error and disturbance of Stern-Gerlach measurements
of a spin-1/2 particle.
We determine the range of the possible values of the error and disturbance for arbitrary 
Stern-Gerlach apparatuses with the orbital degree 
prepared in an arbitrary Gaussian state.
We show that their error-disturbance region is close to the theoretical optimal 
and actually violates Heisenberg's EDR in a broad range of experimental parameters. 
We also show the existence of orbital states in which the error is minimized 
by the screen at a finite distance from the magnet, 
in contrast to the standard assumption.
\end{abstract}

\maketitle

\section{Introduction}

A fundamental feature of quantum measurement is nontrivial 
error-disturbance relations (EDRs), first found by Heisenberg \cite{Hei27}, 
who, using the famous $\gamma$-ray microscope thought experiment, derived the relation 
\begin{equation}
\varepsilon(Q)\eta(P)\geq\frac{\hbar}{2} \label{Hei27}
\end{equation}
between the position measurement error, $\varepsilon (Q)$, and the 
momentum disturbance, $\eta(P)$, thereby caused.
His formal derivation of this relation from the well-established relation
\begin{equation}
\sigma(Q)\sigma(P)\ge\frac{\hbar}{2} \label{Ken27}
\end{equation}
for standard deviations  $\sigma(Q)$ and $\sigma(P)$, due to 
Heisenberg \cite{Hei27} for the minimum uncertainty wave packets 
and Kennard \cite{Ken27} for arbitrary wave functions,
needs an additional assumption on the state change caused by the measurement  \cite{Oza15}.

Nowadays, the state change caused by a measurement is generally described by a completely positive (CP)
instrument, a family of CP maps summing to a 
trace-preserving CP map \cite{Oza84}. 
In such a general description of quantum measurements, Heisenberg's EDR ($\ref{Hei27}$) loses its universal validity, 
as revealed in the debate in the 1980s 
on the sensitivity limit for gravitational wave detection derived by 
Heisenberg's EDR (\ref{Hei27}), but settled questioning the validity of
Heisenberg's EDR  \cite{BVT80,CTDSZ80,Yue83,Cav85,Oza88,Oza89}.
A universally valid error-disturbance relation for arbitrary pairs of observables was derived by one of the authors
 \cite{Oza03a,Oza03b,Oza04} and has recently received considerable attention. 
The validity of this relation, as well as a stronger version of this relation \cite{Bra13, Bra14, Oza14, Oza19}, 
was experimentally tested with neutrons 
 \cite{LW10,ESSGO12,SSEBOH13,DSSOH16} 
and with photons
 \cite{RDMHSS12,BKOE13,WHPWP13,KBOE14,RBB14}.
Other approaches generalizing Heisenberg's original relation $(\ref{Hei27})$
can be found, for example, in 
 \cite{BLW13,BLW14,LYFO14},
apart from the information-theoretic approach 
 \cite{BHOW14,SSDBHOH15}.

Stern-Gerlach measurements \cite{SG22a,SG22b,SG22c} are 
among the most important
quantum measurements, and a number of theoretical analyses 
are available from many authors.
In his famous textbook (see \cite{Boh51}, p. 596), Bohm 
derived the wave function of a spin-$1/2$ particle 
that has passed through the Stern-Gerlach apparatus. 
In his argument, he assumed that the magnetic field points in
the same direction everywhere and varies in strength
linearly 
with the $z$ coordinate of the position as
\begin{equation}\label{BohmMF}
\mathbf{B}
=
\left(
 \begin{array}{c}
 0 \\
 0 \\
 B_0 + B_1z
 \end{array}
\right). 
\end{equation}
However, as Bohm pointed out (see \cite{Boh51}, p. 594), such a magnetic field 
does not satisfy Maxwell's equations.
Theoretical studies \cite{SLB87,BC00,PBC05} of Stern-Gerlach 
measurements with the magnetic field 
\begin{equation}
\mathbf{B}
=
\left(
 \begin{array}{c}
 -B_1x \\
 0 \\
 B_0 + B_1z
 \end{array}
\right)
\end{equation}
satisfying Maxwell's equations were performed only recently. 
According to these studies, if the magnetic field in the center of the beam is sufficiently strong, 
the precession of the spin component 
to be measured becomes small, and hence Bohm's approximation (\ref{BohmMF}) holds. 

Home {\textit et al}. \cite{HPAM11} investigated the error of Stern-Gerlach measurements 
with respect to the distinguishability of apparatus states. 
As an indicator of the operational distinguishability of apparatus states, 
they used the error integral, which is equal to the probability of finding the particle 
in the spin-up state on the lower half of the screen.
They analyzed the error integral 
in the case where
the spin state of the particle just before the measurement is the eigenstate
$\left| \uparrow \right\rangle _z $ of $\sigma _z $ corresponding to the eigenvalue $+1$. 
Nevertheless, the trade-off between the error and disturbance 
in Stern-Gerlach measurements has not been studied in the literature,
even though the subject would elucidate the fundamental limitations of
measurements in quantum theory, as Heisenberg did with the $\gamma$-ray
microscope thought experiment.

In this paper we determine the range of the possible values of the error and 
disturbance for arbitrary Stern-Gerlach apparatuses, 
based on the general theory of the error and disturbance,  
which has recently been developed to establish universally valid reformulations of
Heisenberg's uncertainty relation.
Throughout this paper, we consider an electrically neutral particle with spin $1/2$. 
Following Bohm \cite{Boh51}, 
we assume that the magnetic field of a Stern-Gerlach apparatus is 
represented by Eq.~(\ref{BohmMF}), which is assumed to be
sufficiently strong.
The particle is assumed to stay in the magnet from time 0 to time 
$\Delta t$. 
Only the one-dimensional orbital degree of freedom along the $z$ axis is considered. 
The kinetic energy is not neglected. 
The particle 
having passed through
the magnetic field is assumed to evolve freely from time $\Delta t$ to $\Delta t+\tau$. 
The initial state of the spin of the particle is assumed to be arbitrary. 
The initial state of the orbital degree of freedom is such that mean values 
of the position and momentum are both $0$. 
We study in detail the error $\varepsilon (\sigma _z)$  in measuring $\sigma_z$
with a Stern-Gerlach apparatus and the disturbance $\eta (\sigma _x)$ 
caused thereby on $\sigma _x$ for the orbital degree of freedom to be prepared 
in a Gaussian pure state \cite{Sch86}. We obtain the EDR
\begin{equation}
\left| \frac{\eta (\sigma _x)^2 -2}{2} \right|
\leq
\exp
\left\{
 \left[
  -\mathrm{erf}^{-1}
  \left(
   \frac{\varepsilon (\sigma _z )^{2}-2}{2}
  \right)
 \right]^2
\right\}
\end{equation}
for Stern-Gerlach measurements, where 
$\mathrm{erf}^{-1}$
represents the inverse of the error function 
$\mathrm{erf} (x) = \frac{2}{\sqrt{\pi }}\int _{0}^{x}\exp (-s^2)ds$.
We compare the above EDR with Heisenberg's EDR for spin measurements
\begin{equation}\label{eq:HEDR}
\varepsilon (\sigma _z )^2 \eta (\sigma _z )^2 \geq 1,
\end{equation}
which holds for measurements with statistically independent error and disturbance 
\cite{Oza03a,Oza04}.
We show that Stern-Gerlach measurements violate Heisenberg's EDR in a broad
range of experimental parameters. 
We also compare it with the EDR
\begin{equation}\label{eq:OHEDR}
\left| \frac{\eta (\sigma _x)^2 -2}{2} \right|
\leq  
 1-
 \left(
  \frac{\varepsilon (\sigma _z )^2 -2}{2}
 \right)^2,
\end{equation}
which holds for improperly directed projective measurements
experimentally tested with neutron spin measurements
conducted by Hasegawa and co-workers
\cite{ESSGO12,SSEBOH13}, 
and the tight EDR 
for the range of $(\varepsilon (\sigma _z),\eta (\sigma _x))$ 
values of arbitrary qubit measurements obtained by
Branciard and Ozawa \cite{Bra13,Bra14,Oza14} 
[see \Eq{BOEDRSZ} below].

In Sec.~\ref{measuringprocess} 
the general theory of the error and disturbance is reviewed
and Stern-Gerlach measurements are investigated in the Heisenberg picture in detail. 
In Secs.~\ref{e} and \ref{d} 
the error and disturbance
of Stern-Gerlach measurements are
derived. 
In Sec.~\ref{squeezeded} the EDR  
for Stern-Gerlach measurements is derived. 
In Sec.~\ref{comparisonHome} our research is compared with the previous research conducted by 
Home \textit{et al}.~\cite{HPAM11}.
Sec.~\ref{conclusion} presents a summary.

\section{MEASURING PROCESS \label{measuringprocess}}
For general theory of quantum measurements and their EDRs,
we refer the reader to Appendix A.
\subsection{Spin measurements}
We consider measurements for a spin-1/2 particle $\bS$
and investigate the EDR for the measurements
of the $z$ component, $A=\si_z$, and the disturbance of the $x$ component, 
$B=\si_x$, of the spin.
We suppose that the measurement is carried out by the interaction between
the system $\bS$  prepared in an arbitrary state $\rho$
and the probe $\bP$ prepared in a fixed vector state $\ket{\xi}$
from time 0 to time $\t$ and ends up with the subsequent reading of the meter 
observable $M$ of the probe $\bP$.
We assume the meter $M$ has the same spectrum as the measured observable $\si_z$.
The measuring process, $\bM$, determines the time evolution operator $U$ 
of the composite system of $\bS$ plus $\bP$.
In the Heisenberg picture we have the time evolution of the observables
\beqa
\begin{array}{rclcrclc}
\si_z(0)&=&\si_z\otimes \1, &\quad& \si_z(\t)&=&U^{\da}\si_z(0)U,\\
\si_x(0)&=&\si_x\otimes \1, &\quad& \si_x(\t)&=&U^{\da}\si_x(0)U,\\
M(0)&=&\1\otimes M, &\quad& M(\t)&=&U^{\da}M(0)U.
\end{array}
\eeqa

The /red{probability operator valued measure (POVM)} $\Pi$ of the measuring process $\bM$ is given by
\deq{
\Pi(m)=\av{\xi|P^{M(t_0)}(m)|\xi}.
}
The {\red nonselective operation} $T$ of the measuring process $\bM$ is given by
\deq{
T(\rho)=\Tr_{\cK}[U(\rho\otimes\ketbra{\xi})U^{\dagger}]
}
for any state $\rho$ of $\bS$,
where $\Tr_{\cK}$ is the partial trace over the Hilbert space $\cK$ of the probe $\bP$.

The quantum root-mean-square (rms) error,
$\ep(\si_z)=\ep(\si_z,\bM,\rho)$,
is defined by
\begin{align}
\ep(\si_z)&=
\left(\Tr\left\{\left[M(\t)-\si_z(0)\right]^2\rho\otimes\ketbra{\xi}\right\}\right)^{1/2}.
\end{align}
The quantum rms error $\ep(\si_z)$ has the following properties \cite{Oza19}. 
\begin{itemize}
\item[\textit{(i)}]\textit{Operational definability.}
The quantum rms error $\ep(\si_z)$ is definable by the POVM $\Pi$ of $\bM$
with the observable $\si_z$ to be measured
and the initial state $\RS$ of the measured system $\mathbf{S}$.
\item[\textit{(ii)}]\textit{Correspondence principle.}
In the case where $\si_z(0)$ and $M(t_0)$ commute in 
$\RS\otimes\ketbra{\xi}$, the relation
\begin{equation}
\varepsilon (\si_z) = \varepsilon _{G} (\mu )
\end{equation}
holds for the joint probability distribution 
$\mu$ of $\si_z(0)$ and $M(t_0 )$ in $\RS\otimes\ketbra{\xi}$, where 
$\varepsilon _{G} (\mu )$ is the classical rms error defined by $\mu$.
\item[\textit{(iii)}]\textit{Soundness.}
If $\mathbf{M}$ accurately measures  $\si_z$ in $\RS$, then 
$\varepsilon(\si_z)$ vanishes, i.e., 
$
\varepsilon(\si_z) =0
$.
\item[\textit{(iv)}]\textit{Completeness.}
If  $\varepsilon(\si_z)$ vanishes, then $\mathbf{M}$ accurately measures $\si_z$ in $\RS$.
\end{itemize}
It is known that the completeness property may not hold in the general case \cite{BHL04}, 
but for any dichotomic measurements such that $A(0)^2=M(t_0)^2=\1$ holds
for the measured observable $A$ and the mete observable $M$
as in the case of the present investigation, the completeness property holds \cite{Oza19}.
Thus, the quantum rms error $\ep(\si_z)$ satisfies all the properties 
required for any reliable quantum generalizations of the classical rms 
error, i.e., (i) operational definability, (ii) correspondence principle, 
(iii) soundness,  and (iv) completeness (see Appendix A for further discussions).
 
The quantum rms disturbance $\et(\si_x)=\ep(\si_x,\bM,\rho)$ is defined by
\begin{align}
\et(\si_x)&=
\left(\Tr\left\{\left[\si_x(\t)-\si_x(0)\right]^2\rho\otimes\ketbra{\xi}\right\}\right)^{1/2}.
\end{align}
The quantum rms disturbance  $\et(\si_x)$ has properties analogous to the quantum rms error as follows.  
\begin{itemize}
\item[\textit{(i)}]\textit{Operational definability.}
The quantum rms disturbance $\et(\si_x)$ is definable by the non-selective operation $T$ of $\mathbf{M}$
with the observable $\si_x$ to be disturbed, and the initial state $\RS$ of the measured system $\mathbf{S}$.
\item[\textit{(ii)}]\textit{Correspondence principle.}
In the case where $\si_x(0)$ and $\si_x(t_0)$ commute in $\RS\otimes\ketbra{\xi}$,
the relation
\begin{equation}
\eta(\si_x) = \varepsilon _{G} (\mu )
\end{equation}
holds for the joint probability distribution $\mu$ of $\si_x(0)$ and $\si_x(t_0 )$ in $\RS\otimes\ketbra{\xi}$.
\item[\textit{(iii)}]\textit{Soundness.}
If $\mathbf{M}$ does not disturb $\si_x$ in $\RS$, then 
$\eta(\si_x)$ vanishes.
\item[\textit{(iv)}]\textit{Completeness.}
If  $\eta(\si_x)$ vanishes, then $\mathbf{M}$ does not disturb $\si_x$ in $\RS$.
\end{itemize}
It is known that the completeness property may not hold in the general case
(see \cite{06QPC}, p. 750), 
but for any dichotomic observables such that $B^2=\1$ to be disturbed
as in the case of the present investigation the completeness property always holds \cite{Oza19}.
Thus, the quantum rms disturbance $\et(\si_x)$ satisfies all the properties 
required for any reliable quantum generalizations of the classical rms change of
observable $B$ from time 0 to $\t$, i.e., 
(i) operational definability, (ii) correspondence principle, 
(iii) soundness,  and (iv) completeness (see Appendix A for further discussion). 

\begin{figure}[htb]
\begin{center}
\includegraphics[width=0.45\textwidth]{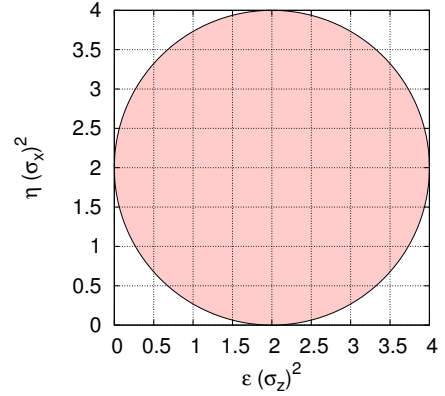}
\caption{$\varepsilon (\sigma _z) ^2$ \!-\! $\eta (\sigma _x) ^2$ plot of tight EDR
 \eq{BOEDRS} for spin measurements in the state satisfying \Eq{condition0}.}
\label{borel}
\end{center}
\end{figure}

Since $\si_z^2=\si_x^2=\1$ and $M^2 = \1$, 
from \Eq{BEDRM} we obtain 
\begin{equation}
\hat \varepsilon (\si_z)^2  + \hat \eta (\si_x)^2+ 2  \hat \varepsilon(\si_z) \hat \eta(\si_x) \sqrt{1-D_{\si_z\si_x}^2}
\geq D_{\si_z\si_x}^2,
\label{eq:BOEDRS}
\end{equation}
where
\begin{align}
D_{\si_z\si_x}&=\mathrm{Tr}(\left| \sqrt{\RS }\si_y \sqrt{\RS }\right|),\\
\hat \varepsilon (\si_z)&=\sqrt{
 1-
 \left(
  \frac{\varepsilon (\sigma _z )^2 -2}{2}
 \right)^2
},\\
\hat \eta (\si_x)&=\sqrt{
 1-
 \left(
  \frac{\eta (\sigma _x )^2 -2}{2}
 \right)^2
}
\end{align}
from the EDR obtained by Branciard \cite{Bra13} for pure states  
and extended to mixed states by Ozawa \cite{Oza14}.
In the case where
\begin{equation}
\left\langle \sigma _z \right\rangle _{\RS } = \left\langle \sigma _x \right\rangle _{\RS } = 0, 
\label{eq:condition0}
\end{equation}
\Eq{BOEDRS} is reduced to the tight relation
\begin{equation}
\left[
 \varepsilon  (\sigma _z) ^2 -2
\right]^2
+
\left[
 \eta (\sigma _x) ^2 -2
\right]^2
\leq 4
\label{eq:BOEDRSZ}
\end{equation}
as depicted in Fig. \ref{borel}.

Lund and Wiseman \cite{LW10} proposed a measurement model 
measuring the Pauli $\si_z=\ketbra{0}-\ketbra{1}$ observable of an abstract qubit
described by the Hilbert space $\cH=\C^2$ with a computational basis 
$\{\ket{0},\ket{1}\}$.
The probe is another qubit prepared in the state 
$\ket{\xi(\theta)}=\cos\theta\ket{0}+\sin\theta\ket{1}$ and the meter observable $M$
is chosen as the Pauli $\si_z$ observable of the probe.
The measuring interaction is described by the unitary operator $U_{\text{CNOT}}$ 
on $\C^2\otimes\C^2$ performing the controlled-NOT operation 
controlled on the measured qubit.
Thus,  the measuring process is specified as 
$\M(\theta)=(\C^2,\ket{\xi(\theta)},U_{\text{CNOT}},\si_z )$.
Then, for the system state $\ket{\psi}=\ket{\si_y=+1}
=(1/\sqrt{2})(\ket{0}+i\ket{1})$, 
which satisfies condition \eq{condition0} for $\rho=\ketbra{\psi}$,
the measurement error $\ep(\si_z)$ 
of $\M(\theta)$
for $A=\si_z$ 
and the disturbance $\et(\si_x)$ of $\M(\theta)$ 
for $B=\si_x$ is given by 
\begin{align}
\ep(\si_z)&=2|\sin \theta|,\\
\et(\si_x)&=\sqrt{2}|\cos\theta-\sin\theta|.
\end{align} 
Thus, the error $\ep(\si_z)$ and disturbance $\et(\si_x)$ satisfy the relation
\begin{align}
\left[\ep(\si_z)^2-2\right]^2+\left[\et(\si_x)^2-2\right]^2=4,
\end{align}
and attain the bound for the EDR \eq{BOEDRS}.
Experimental realizations of this EDR for optical 
polarization measurements were reported by Rozema \textit{et al.}~\cite{RDMHSS12}
and others \cite{BKOE13,RBB14,KBOE14}.   

In this paper we consider another type of measurement model,
known as Stern-Gerlach measurements,
measuring the $z$ component of the spin of a spin-1/2 particle,
and investigate the admissible region of the error and disturbance
obtained from Gaussian orbital states.  

\subsection{Stern-Gerlach measurements}\label{se:S-GM}
\begin{figure}[h]
\begin{center}
\includegraphics[width=0.45\textwidth]{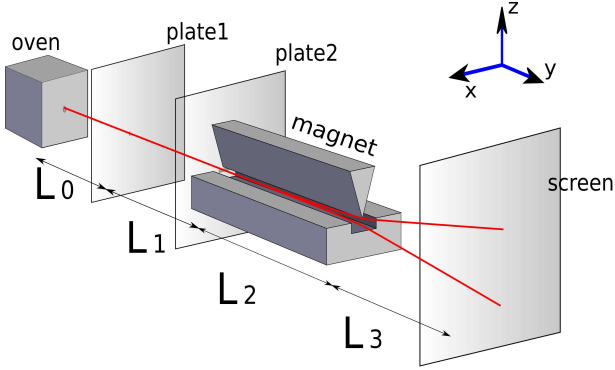}
\end{center}
    \caption{
    Illustration of the experimental setup for a  Stern-Gerlach measurement. 
    The relations between the length and the time interval are $L_2 = v_y \Delta t$ 
    and $L_3 = v_y \tau $.
    }
    \label{sgexp}
\end{figure}

Let us consider a measurement of the spin component of an electrically neutral spin-1/2 particle 
with a Stern-Gerlach apparatus.
A particle moving along the $y$ axis passes through an inhomogeneous 
magnetic field and then the orbit is deflected, 
depending on the spin component of the particle along the direction of the magnetic field. 
This situation is illustrated in Fig.~\ref{sgexp}.
To analyze this measurement, we make the following assumptions.
\begin{itemize}
\item[(i)]
The magnetic field points everywhere on the $z$ axis. 
\item[(ii)]
The strength of the magnetic field increases proportional to the $z$-coordinate,
\begin{equation}
B_z = B_0 + B_1 z,
\end{equation}
where $B_0$ and $B_1$ are real numbers representing the value at the origin and the gradient of $B_z$, 
	 respectively. 
\item[(iii)]
The velocity, $v_y$, in the $y$-direction is large in comparison  
with the motion in the $x$-$z$ plane and the length $L_2$ is  
large in comparison with the separation of the pole faces.
Thus we can treat the times $\De t=L_2/v_y$ and $\tau=L_3/v_y$ as deterministic
for our purpose, because the determination of the spin does not depend in a sensible 
way on the precise evaluation of $\Delta t$ and $\tau$  (see \cite{Boh51}, pp. 595-596).
\end{itemize}

To describe the measuring process $\bM$ of a Stern-Gerlach measurement, 
the measured system $\bS$ is taken as the spin degree of freedom described by 
the two-dimensional state space $\cH$ with the Pauli operators
$\si_x$, $\si_y$, and $\si_z$ describing the $x$, $y$, and $z$ components of the spin, respectively, of the
spin-1/2 particle.
The probe system $\bP$ is taken as the orbital degree of freedom in the $z$ direction
described by the Hilbert space, $\cK$, of wave functions with position $Z$
and momentum $P$ satisfying the canonical commutation relation
\beq
[Z,P]=i\hbar.
\eeq
The particle enters the magnetic field at time 0, emerges out of the 
magnetic field at time $\De t$, and freely evolves until time $\De t+\ta$
at which the particle reaches the screen and the observer can measure the
meter observable $M$ that assigns $+1$ or $-1$ 
depending on the particle $z$ coordinate, $Z$, as $M=f(Z)$, 
with the function $f$ such that
\begin{equation}
f(z)=
\begin{cases}
-1& \mbox{if  $z\geq 0$} \\
+1& \mbox{otherwise}. 
\end{cases}
\end{equation}
Thus, the measuring process starts at time 0, when the system $\bS$ is in any input state
$\rho$ and the probe $\bP$ is prepared in the fixed state $\kxi$,
and ends up at time $\t=\De t+\ta$.
The time evolution operator $U=U(\De t+\ta)$ of the composite system $\bS+\bP$ during
the measurement is determined by the time-dependent
Hamiltonian $H(t)$ of the particle given by
\begin{widetext}

\begin{equation}
H(t)\!=\!\left\{
\begin{array}{lr}
\mu\sigma _z\otimes\left( B_0 + B_1 Z\right) 
 +\displaystyle\frac{1}{2m}\1\otimes P^2, 
 &
\hspace{-17pt} 
0 \leq t \leq \Delta t \\
 \displaystyle\frac{1}{2m}\1\otimes P^2, 
& 
 \hspace{-17pt} 
\Delta t \leq t \leq \Delta t + \tau,
\end{array}
 \right.
\end{equation}
where $\mu $ denotes the magnetic moment of the particle 
and $m$ denotes the mass of the particle. 
By solving the Schr\"{o}dinger equation, we obtain
 the time evolution operator $U(t)$ of $\bS+\bP$ for $0\le t\le \De t+\ta$
 by
\begin{align}
U(t)
=
\left\{
\begin{array}{lr}
\exp
\left\{
 \displaystyle\frac{t}{i\hbar }
 \left[
  \mu \sigma _z
  \otimes
  \left(
    B_0 \!+\! B_1 Z
  \right)
  +
  \displaystyle\frac{1}{2m} 
  \1 \otimes P^2
 \right]
\right\},
& 0 \leq t \leq \Delta t \\
\exp \left(\displaystyle\frac{t - \Delta t}{2i\hbar m} \1 \otimes P^2 \right) 
\exp
\left\{
  \displaystyle\frac{\Delta t}{i\hbar }
 \left[
  \mu \sigma _z \!
  \otimes
  \!
  \left(
   B_0  \! +\! B_1 Z
  \right)
  +
  \displaystyle\frac{1}{2m} \1 \otimes P^2
 \right]
\right\},
& \Delta t \leq t \leq \Delta t + \tau.
\end{array}
\right.
\end{align}

\end{widetext}
To describe the time evolution of the composite system $\bS+\bP$
in the Heisenberg picture, we introduce Heisenberg operators for $0\le t\le \De t+\ta$
as 
\begin{align}
Z(0)=\1 \otimes Z, &\quad Z(t)=U(t)^{\da}Z(0)U(t), \\
P(0)=\1 \otimes P, &\quad P(t)=U(t)^{\da}P(0)U(t), \\
\si_j(0)=\si_j\otimes \1, &\quad \si_j(t)=U(t)^{\da}\si_j(0)U(t),
\end{align}
where $j=x,y,z$.
For the relation between the time evolution operators in 
the Heisenberg picture and the Schr\"{o}dinger picture, 
we refer the reader to Appendix D.
 
We also use the matrix representations of Pauli operators as
\begin{align} 
\sigma _x
=
\left(
\begin{array}{cc}
0&1\\
1&0
\end{array}
\right),
& \quad
\sigma _y
=
\left(\begin{array}{cc}
0&-i\\
i&0
\end{array}
\right), \quad
\sigma _z =
\left(
\begin{array}{cc}
1&0\\
0&-1
\end{array}
\right).
\end{align}

By solving Heisenberg equations of motion for 
$Z(t)$, $P(t)$, $\si_x(t)$, $\si_y(t)$, and $\si_z(t)$, 
as shown in Appendix \ref{se:App_SHEM},
we have
\begin{align}
Z(\Delta t + \tau )&=
Z(0)
+
\frac{\Delta t + \tau}{m}P(0)\notag\\
&\quad  -
\frac{\mu B_1 \Delta t}{m}
\left(
 \tau + \frac{\Delta t }{2}
\right)
\sigma _z(0), \label{Zjustaftermeas}\\
P(\Delta t + \tau )
&=
P(0) - \mu B_1 \Delta t \sigma _z (0), \\
\si_x(\Delta t + \tau )&=\left(
 \begin{array}{cc}
  0        & \exp \left[ iS(\De t) \right] \\
  \exp \left[ -iS(\De t) \right] & 0
 \end{array}
\right), \label{sigma_zjustaftermeas}\\
\si_y(\Delta t + \tau )
&=\left(
 \begin{array}{cc}
  0          & -i \exp \left[ iS(\De t) \right] \\
  i \exp \left[ -iS(\De t) \right] & 0
 \end{array}
\right),\\
\sigma _z (\Delta t + \tau )
&=
\sigma _z (0),
\end{align}
where
\begin{eqnarray}
S(\De t)&=&
\frac{2 \mu \De t }{\hbar }
\left[
 B_0 + B_1 \left( Z(0) + \frac{\De t}{2m}P (0)\right)
\right].
\end{eqnarray}

\section{Error \label{e}}
Let us consider the quantum rms error of a Stern-Gerlach measurement $\bM$ 
of the $z$ component $\si_z(0)$ of the spin at time 0 using the meter observable 
\deq{M(\De t+\ta)=f(Z(\De t+\ta)),} introduced in Sec.~\ref{measuringprocess}.
The noise operator $N$ of this measurement is given by
\begin{equation}
N = M (\Delta t + \tau) - \sigma _z (0).
\end{equation}

The initial state $\rho$ of the spin $\bS$ is supposed to be an arbitrary state
with the matrix
\beql{Bloch}
\rho
= \frac{1}{2}(\1+n_x\si_x+n_y\si_y+n_z\si_z)
\eeq
where $n_x,n_y,n_z\in \R$ and $n_x^2+n_y^2+n_z^2\le 1$,
so that the initial state of the composite system $\bS+\bP$ is
given by
$\rho \otimes \left| \xi \right\rangle \left\langle \xi \right|$,
where $\ket{\xi}$ is a fixed but arbitrary wave function describing the
initial state of the orbital degree of freedom $\bP$.
Then the error, namely, the quantum rms error, of this measurement of $\si_z$ is given by
\begin{equation}
\varepsilon (\sigma _z) = \sqrt{\langle N^2 \rangle _{\rho \otimes \left| \xi \right\rangle \left\langle \xi \right| } }, 
\end{equation}
where we abbreviate
$
\mathrm{Tr}(A \rho )
$
as
$
\langle A \rangle_{\rho }
$
for observable $A$ and density operator $\rho $.
We will give an explicit formula for $\varepsilon (\sigma _z)$,
which eventually shows that the error depends only on the parameter $n_z$ in \Eq{Bloch}. 

Let
\begin{align}
U_t& = \exp \left[ \frac{t}{2i\hbar m}P^2 \right],\\
\tilde{U}_t& = \1_{\bS}  \otimes U_t,\\
g_0 &=
\frac{\mu B_1 \Delta t }{m}
\left(
 \tau +\frac{\Delta t}{2}
\right). 
\end{align}
From Eq.~(\ref{Zjustaftermeas}) we have
\begin{align}
\lefteqn{Z(\Delta t + \tau ) }\quad\nn\\
&=
\tilde{U}_{\De t+\ta}^{\dag}
\bmat
Z-g_0& 0\\
0&Z+ g_0
\emat 
\tilde{U}_{\De t+\ta}.
\end{align}
Thus, we have
\begin{align}\label{eq:noise-operator}
N
&=f(Z(\Delta t + \tau ))-\si_z(0) \nonumber \\
&=
2\tilde{U}_{\De t+\ta}^{\dag}
\bmat
-\chi _+(Z-g_0)& 0\\
0& \chi _-(Z+ g_0)\\
\emat
\tilde{U}_{\De t+\ta},
\end{align}
where
\begin{align}
\chi _+ (z) &= 
\begin{cases}
 1 & \mbox{if $z \geq 0$}, \\
 0 & \mbox{otherwise},
\end{cases}\\
\chi  _- (z)&= 1-\chi _+(z),\\
 f(z)&=1-2\chi_{+}(z).
\end{align}
It follows that 
\begin{align}
N^2
&=
4\tilde{U}_{\De t+\ta}^{\dag}
\bmat
\chi _+(Z-g_0)& 0\\
0& \chi _-(Z+ g_0)\\
\emat
\tilde{U}_{\De t+\ta}.
\end{align}
Therefore, we have
\begin{align}
\varepsilon (\sigma _z) ^2 
&=
\langle N^2 \rangle _{\rho \otimes \left| \xi \right\rangle \left\langle \xi \right|} \notag \\
&=
\left\langle 
\xi| \Tr_{\bS}[N^2 \rho]|\xi
\right\rangle\nn\\
&=
2(1+n_z)\langle 
\xi|U_{\De t+\ta}^{\dag}\chi _+(Z-g_0)U_{\De t+\ta}^{\dag}|\xi\rangle\nn\\
& \quad
+2(1-n_z)\langle \xi|U_{\De t+\ta}^{\dag}\chi _-(Z-g_0)U_{\De t+\ta}^{\dag}|\xi\rangle.
\end{align}
Consequently, we have
\begin{align}
\varepsilon (\sigma _z)^2
&=
2(1+n_z)
\int _{g_0}^{\infty }
|U_{\De t+\ta}\xi (z)|^2 dz \biggr. \nonumber \\
\biggl. & \qquad \qquad \qquad
+
2(1-n_z)
\int _{-\infty}^{-g_0}
|U_{\De t+\ta}\xi (z)|^2 dz. \biggr. \label{twovaluederror}
\end{align}

\section{Disturbance \label{d}}
Let us consider the quantum rms disturbance, $\eta (\sigma _x)$, 
for the $x$-component of the spin in Stern-Gerlach measurements. 
The disturbance operator, $\sigma _x $, is given by
\begin{equation}
D = \sigma _x (\Delta t + \tau )-\sigma _x(0).
\end{equation}
From Eq.~(\ref{sigma_zjustaftermeas}) we have
\begin{equation}
D =
\left(
 \begin{array}{cc}
 0 & \exp \left[ iS(\Delta t ) \right] -1 \\
 \exp \left[ -iS(\Delta t ) \right] -1 & 0
 \end{array}
\right).
\end{equation}
Consequently, we have
\begin{equation}
D^2 =
\1 \otimes 
\left[
 2 - 2 \cos S(\Delta t )
\right],
\end{equation}
and thus
\begin{align}
&\eta (\sigma _x) ^2 \nonumber \\
&=
2 \!
- \!
2
\left\langle
 \cos \!
 \left\{ \!
  \frac{2 \mu \Delta t }{\hbar }
  \left[ \!
   B_0 + B_1
   \left( \!
    Z + \frac{\Delta t}{2m}P 
   \right) \!
  \right] \!
 \right\}
\right\rangle _{\xi }. \label{disturbance}
\end{align}

\section{Error and disturbance for Gaussian states \label{squeezeded}}
Let us consider the error and disturbance in Stern-Gerlach measurements under the condition that the 
orbital state of the particle is in the family $\cG$ of Gaussian states given by 
\begin{equation}
\mathcal{G}
=
\left\{
 \xi _{\lambda }\in \mathit{L}^2 (\mathbb{R})
 \mathrel{} \left| \mathrel{}
 \begin{array}{l}
 \xi _{\lambda } (z) = A \exp (-\lambda z^2) \\[0.5em]
 \displaystyle \int _{-\infty}^{\infty}|\xi _{\lambda } (z)|^2 dz =1 \\[1em]
 \lambda \in \mathbb{C}, 
 \Re(\lambda ) >0  
 \end{array}\right.
\right\}.
\end{equation}
This family of states consists of all Gaussian pure states \cite{Sch86},
 whose mean values of the position and momentum are both $0$. 
For simplicity, it is assumed that the spin state of the particle is in the eigenstate of the spin component
$\sigma _y$.
It is easy to minimize the error of the measurement with respect to the mean values of the position and
momentum. 
In particular, $\mathcal{G} $ is the family of optimal states for the measurement among the Gaussian pure states if the 
spin state of the particle is the
 eigenstate of $\sigma _y$. 
We remark that the equality in the Schr\"{o}dinger inequality 
[see Eq.~(\ref{schr_ineq}) ] holds for any 
state $\xi $ in $\mathcal{G}$, i.e.,
\begin{equation}
\langle Z^2 \rangle _{\xi}
\langle P^2 \rangle _{\xi}
-
\frac{1}{4}\langle\{Z, P\} \rangle _{\xi } ^2
=
\frac{\hbar ^2}{4}. \label{eqschrodingerineq}
\end{equation}
Here we use the abbreviation
$\langle A \rangle _{\xi } = \av{\xi| A |\xi}$. 
The converse also holds, that is, any state $\xi $ satisfying 
$\langle P \rangle _{\xi } = \langle Z \rangle _{\xi } = 0 $ and Eq.~(\ref{eqschrodingerineq}) 
belongs to $\mathcal{G}$.

Let us consider the range of the error and disturbance of Stern-Gerlach measurements.
Let
\begin{align}
V(\ps,t)&=
\left\langle
 \left(
  Z + \frac{t}{m}P
 \right)^2
\right\rangle_{\ps}
\end{align}
for any orbital state $\psi$.
For the disturbance $\eta (\sigma _x)$, from Eq.~(\ref{disturbance}) we have
\begin{align}
& \eta (\sigma _x) ^2 \nonumber \\
&=
2 
-
2
\left\langle
 \cos
 \left[
  \frac{2 \mu \Delta t }{\hbar }
   \left(
    B_0 + B_1Z
   \right)
  \right]
\right\rangle _{U_{\De t/2}\xi _{\lambda } } \nonumber \\ 
&=
2-
\frac{2}{\sqrt{2 \pi V(\xi_{\la},\De t/2)}} \nonumber \\ 
 \times & 
\int_{-\infty}^{\infty}
\exp
\left(
 -\frac{z^2}
 {2V(\xi_{\la},\De t/2) }
\right)
\cos
\left[
 \frac{2 \mu \Delta t }{\hbar }
 \left(
  B_0 + B_1 z
 \right)
\right] dz \nonumber \\
&=
2-2
\exp \!
\left(
 -\frac{2 \mu ^2 \! B_1^2 \! \Delta t^2}{\hbar ^2}V(\xi_{\la},\De t/2)
\right)
\cos
 \frac{2 \mu \Delta t B_0}{\hbar }.
\end{align}
From the above formula, the disturbance is determined by $V(\xi_{\la},\De t/2)$ and the parameters of the magnet
if the orbital state is in $\mathcal{G}$.
Now, for a fixed constant $v$ let us find the error for state $\xi _{\lambda } $ in $\mathcal{G}$ and 
time interval 
$\Delta t$ satisfying $V(\xi_{\la},\De t/2)=v$. 
In the following, we fix the time interval $\Delta t $.

From Eq.~(\ref{twovaluederror}) we have
\begin{eqnarray}
\varepsilon (\sigma _z) ^2
&=&
4 \int_{g_0}^{\infty }
\left|
U_{\De t+\ta}\xi _{\lambda } (z)
\right|^2
dz \nonumber \\
&=&
\frac{4}{\sqrt{\pi}}
\int_{{g_0}/{\sqrt{2V(\xi_{\la},\De t+\ta)}}}^{\infty }
\exp(-w^2)dw. \label{errorforGaussianpurestates}
\end{eqnarray}
Here we use the relation $n_z=0$, which is obtained from the assumption 
that the mean value of the $z$ component of the spin of 
the particle
is $0$.
Equation (\ref{errorforGaussianpurestates}) shows that the error is minimized 
by maximizing the lower limit of the integration
$
g_0 /\sqrt{2V(\xi_{\la},\De t+\ta) }
$. 
First, we fix the state $\xi _{\lambda }$ and focus on the time interval $\tau $.
Let
$W_{\xi_{\lambda }}(\tau )=g_0/\sqrt{2V(\xi_\la,\Delta t+\tau) }$.
	From now on, we suppose $B_1 \leq 0$.
As shown in Appendix \ref{se:App_Sup}, if
\begin{equation}
m
\left\langle
 \left\{
  Z, P
 \right\}
\right\rangle_{\xi _{\lambda } }
+
\left\langle
 P^2
\right\rangle_{\xi _{\lambda } }
\Delta t
<0 \label{condition}
\end{equation}
holds, then $W_{\xi_{\lambda} } (\tau ) $ assumes the maximum value 
\begin{equation}
W_{\xi_{\lambda}}(\tau _0)
=
\frac
{\sqrt{2V(\xi_\la,\Delta t /2)}\mu B_1 \Delta t}{\hbar}
\end{equation}
at
\begin{align}
\tau =& \tau _0 \notag \\
=&
-\frac
{
 4m^2
 \left\langle
  Z^2
 \right\rangle _{\xi _{\lambda } }
 +
 3m
 \left\langle
  \left\{
   Z,  P
  \right\}
 \right\rangle_{\xi _{\lambda } }
 \Delta t
 +
 2 \left\langle P^2 \right\rangle_{\xi _{\lambda }} \Delta t ^2
}
{
 2
 \left(
  m
  \left\langle
   \left\{
    Z, P
   \right\}
  \right\rangle_{\xi _{\lambda } }
  +
  \left\langle
   P^2
  \right\rangle_{\xi _{\lambda } }
  \Delta t
 \right)}.
\end{align}
On the other hand, if  condition (\ref{condition}) does not hold, 
the supremum of $W_{\xi_{\lambda}}(\tau )$ is given by
\begin{equation}
\sup_{\tau \geq 0} W_{\xi_{\lambda} }(\tau )=\lim_{\tau \to \infty } W_{\xi _{\lambda }}(\tau )
=
\frac{\mu B_1 \Delta t}{\sqrt{2}}
\left\langle
 P^2
\right\rangle _{\xi _{\lambda }} ^{-1/2}. 
\end{equation}

Now let us consider the maximization of $W_{\xi _{\lambda }}(\tau )$ with respect to the state 
$\xi _{\lambda }$.
For any pair of orbital states $\psi $ and $\phi $ in $\mathcal{G}$ satisfying 
$V(\psi, \Delta t/2)=v$ and $V(\phi, \Delta t/2)=v$, respectively,
if $\psi $ satisfies condition (\ref{condition}), then
\begin{equation}
W_{\psi }(\tau _0) \geq \lim_{\tau \to \infty }W_{\phi }(\tau )
\end{equation}
holds, since $W_{\psi }(\tau _0)/\lim_{\tau \to \infty }W_{\phi }(\tau )\ge 1$ by
the Kennard inequality (\ref{Ken27}).
Therefore, we obtain the supremum of $W_{\xi _{\lambda }}(\tau)$ with respect to the
state $\xi _{\lambda }$ and time interval $\tau $ as
\begin{equation}
\sup _{\mathrm{Re}(\lambda)>0, \tau \geq 0} W_{\xi _{\lambda }}(\tau )
=
\frac{\sqrt{2v}\mu B_1 \Delta t}{\hbar }.
\end{equation}
See Appendix \ref{se:App_Sup} for the detailed derivation.

Although the above argument is for finding the range of the error and disturbance that 
Stern-Gerlach measurements can assume, it contains one more 
important assertion.  
That is, the calculation suggests that the error of Stern-Gerlach measurements is minimized 
by placing the screen at a finite distance from 
the magnet under the condition represented by (\ref{condition}), in contrast to the conventional 
assumption that the error is minimized by 
placing the screen at infinity. 
If a state in $\mathcal{G}$ satisfies condition (\ref{condition}), then the correlation term
\cite{Yue83}
$
\left\langle
 \left\{
  Z -\left\langle Z \right\rangle _{\xi _{\lambda }},
  P -\left\langle P \right\rangle _{\xi _{\lambda }}
 \right\}
\right\rangle_{\xi _{\lambda } }
$
is negative, and this leads to a narrowing of the standard deviation of the position of the particle during the
free evolution (see Appendix \ref{se:contractive}).
Such a class of states was introduced by Yuen  \cite{Yue83}
and they are known as contractive states.

Let us return to the problem of finding the range of values of the error and disturbance that Stern-Gerlach measurements can assume.
Now setting $W_0= {\sqrt{2v}\mu B_1 \Delta t}/{\hbar }$, the disturbance and the infimum of the error under the condition that
$V(\lambda, \Delta t/2)=v $ for fixed $\Delta t$ and $v$ are
\begin{eqnarray}
\eta(\si _x) ^2
&=&
2-2
\exp
\left(
 -W_0^2
\right)
\cos
 \frac{2 \mu \Delta t B_0}{\hbar}, \\
\inf _{\lambda, T}
\varepsilon(\si_z) ^2
&=&
\frac{4}{\sqrt{\pi}}
\int_{W_0}^{\infty }
\exp(-w^2)dw,
\end{eqnarray}
respectively. 
By varying the parameter of the magnet $B_0$, we obtain the range of the disturbance as
\begin{equation}
2-2
\exp
\left(
 -W_0^2
\right)
\leq
\eta (\sigma _x)^2
\leq
2+2
\exp
\left(
 -W_0^2
\right).
\end{equation}
We obtain the range of the disturbance and the infimum of the error of Stern-Gerlach measurements
for each constant $v$.
By varying $v$, we obtain the range of the error and disturbance as the 
inequalities
\begin{equation}
\left| \frac{\eta (\sigma _x)^2 - 2}{2} \right|
\leq
\exp 
\left\{-\left[\mathrm{erf}^{-1}\left(\frac{\varepsilon (\sigma _z)^2-2}{2}\right)\right] ^2\right\}, 
\end{equation}
\begin{equation}
0 \leq \varepsilon (\sigma _z)^2 \leq 2, 
\end{equation}
where 
$\mathrm{erf}^{-1}$
represents the inverse of the error function 
$\mathrm{erf} (x) = ({2}/{\sqrt{\pi }})\int _{0}^{x}\exp (-s^2)ds$.
The square of the error varies from $0$ to $ 2$ since $W_0$ is positive.

We now remove the constraint $B_1 \leq 0$.  
For $B_1 \geq 0$, similarly to the above discussion, we have
\begin{equation}
\left| \frac{\eta (\sigma _x)^2 - 2}{2} \right|
\leq
\exp 
\left\{-\left[\mathrm{erf}^{-1}\left(\frac{\varepsilon (\sigma _z)^2-2}{2}\right)\right] ^2\right\}, 
\end{equation}
\begin{equation}
2 \leq \varepsilon (\sigma _z)^2 \leq 4.
\end{equation}
Therefore, we have
\begin{equation}\label{eq:edSG}
\left| \frac{\eta (\sigma _x) ^2 -2}{2} \right|
\leq  
\exp
\left\{
 -
 \left[
  \mathrm{erf}^{-1}
  \left(
   \frac{\varepsilon (\sigma _z )^2-2}{2}
  \right)
 \right]^2
\right\} .
\end{equation}
The plot of this region is shown in Fig.~\ref{figed}.

For comparison, the figure shows the plot of the boundary of the Branciard-Ozawa
tight EDR \eq{BOEDRSZ} for 
general spin measurement.
From this plot, we conclude that the range of the error and disturbance for Stern-Gerlach measurements
considered in this paper is close to the theoretical optimal 
given by the Branciard-Ozawa tight EDR (\ref{eq:BOEDRSZ}).
Here the range of the error and disturbance for Stern-Gerlach measurements is also compared with
Heisenberg's EDR \eq{HEDR} (green line)
and the EDR (\ref{eq:OHEDR}) for the neutron experiment
\cite{ESSGO12,SSEBOH13} (black line).
We conclude that Stern-Gerlach measurements actually violate Heisenberg's EDR (\ref{eq:HEDR}) 
in a broad range of experimental parameters. 

Roughly speaking, the parameter $v$ represents the spread of the wave packet of the particle 
in the Stern--Gerlach magnet. The reason why $v$ appears in the formula of the disturbance 
is that the particle in the Stern-Gerlach magnet is exposed to the inhomogeneous magnetic field 
and its spin is precessed in an uncontrollable way. This uncontrollable precession occurs because the 
position of the particle is uncertain while the magnetic field is inhomogeneous and hence 
depends on the position. The disturbance of the spin along the $x$ axis is caused by 
this uncontrollable precession around the $z$ axis. This is why $v$ appears in the formula of the disturbance. 
On the other hand, the error in our Stern-Gerlach setup comes from the non zero dispersion of the 
$z$ component of the particle position when the particle has reached the screen.
The smaller 
the dispersion of the particle position when the particle has reached the screen,
the greater the dispersion of the $z$ component of the particle position in the Stern-Gerlach magnet.
This is why $v$ appears in the formula of the error.

\begin{widetext}

\begin{figure}[h]
\begin{center}
\includegraphics[width=0.9\textwidth]{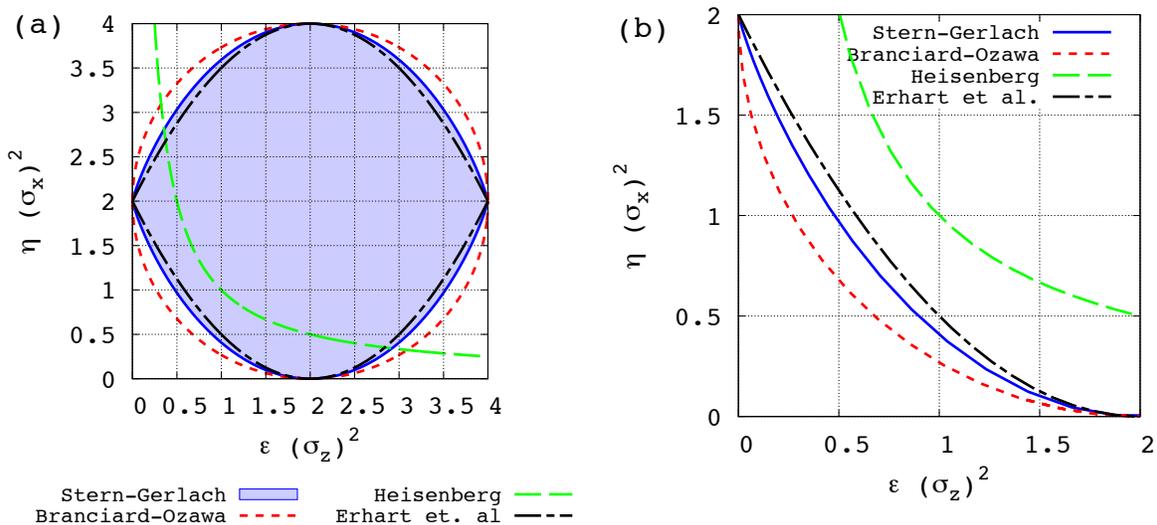}
\caption{
(a) Range of error and disturbance for Stern-Gerlach measurements.
The blue region is the region \eq{edSG} that Stern-Gerlach measurements can achieve.
The red dotted line is the boundary of the Branciard-Ozawa tight EDR (\ref{eq:BOEDRSZ}).
The green dashed line is the boundary of Heisenberg's EDR (\ref{eq:HEDR}).
The black dash-dotted line is the theoretical boundary \eq{OHEDR} of the EDR of the experiment 
conducted by Erhart
and co-workers \cite{ESSGO12,SSEBOH13}.
The error-disturbance region of Stern-Gerlach measurements is close to the theoretical optimum 
given by the Branciard-Ozawa tight EDR (\ref{eq:BOEDRSZ})
and actually violates Heisenberg's EDR (\ref{eq:HEDR}) in a broad range of experimental parameters. 
(b) Enlarged plot for the part $[0,2]\times[0,2]$.
}
\label{figed}
\end{center}
\end{figure}

\vspace{20pt}
\end{widetext}

\section{Comparison with ``aspects of nonideal Stern-Gerlach experiment and testable ramifications"} \label{comparisonHome}
Home \textit{et al.}~\cite{HPAM11} discussed the same error of Stern-Gerlach measurements as our paper
 does for similar conditions.
We consider in what sense their paper is related to ours and we compare its results with ours.
They derived the wave function of a particle in the Stern-Gerlach apparatus under the following 
conditions.
\begin{itemize}
\item[(i)]
The magnetic field is oriented along the $z$ axis everywhere and the gradient of the $z$ component of the magnetic 
field is non zero only in the $z$ direction. 
\item[(ii)]
The initial orbital state is a Gaussian state whose mean values of the position and momentum, and the correlation term 
of the particle in the wave function 
are all zero.
\item[(iii)]
Unlike Bohm's discussion \cite{Boh51}, the kinetic energy of the particle in the magnetic field is not neglected.
\end{itemize}
Based on their argument, they discussed the distinguishability of the value of the measured observable by observing the probe system directly
in Stern-Gerlach measurements. 
To consider this problem, they introduced the two indices,
\begin{eqnarray}
&&I
:=
\left|
 \int _{-\infty}^{\infty} \int _{-\infty}^{\infty} \int _{-\infty}^{\infty}
 \psi _+ ^* (\mathbf{x}, \tau) \psi _- (\mathbf{x}, \tau) d\mathbf{x}
\right|,
\\ 
&&E(t)
:=
\int _{-\infty}^{0} \int _{-\infty}^{\infty} \int _{-\infty}^{\infty}
\left| \psi _+ (\mathbf{x}, t) \right| ^2 dx dy dz,
\end{eqnarray}
where $\psi _{\pm }$ are the wave functions of the particle in the Schr\"{o}dinger picture 
whose spin $z$ components are $\pm 1/2$, respectively.
The origin of time is taken to be the moment when the particle enters the Stern-Gerlach magnet.
In addition, $\tau $ is the time at which the particle emerges from the Stern-Gerlach magnet 
($\tau $ corresponds to $\Delta t$ in our notation) 
and $t$ is any time after emerging from the Stern-Gerlach magnet ($t$ corresponds to $\Delta t + \tau$ 
in our notation).
Namely, they adopted the inner product $I$ of the two wave functions with different spin 
directions, 
and the probability $E(t)$ of finding the particle with the spin $z$ components of 
$+ 1/2$ and $-1/2$ within the lower and upper half planes, respectively, at time $t$.
They concluded that $I$ always vanishes whenever $E(t)$ vanishes, but that $E(t)$ does not necessarily vanish even 
when $I$ vanishes.  

We discuss the relation between their paper and ours.
The relation between the quantities $E(t)$ and $\varepsilon (\sigma _z) $ is
\begin{equation}
\varepsilon (\sigma _z)^2 =4 E(t).
\end{equation}
Although this relation is model dependent, it bridges the two approaches and will enforce a theoretical 
background for our definition of a sound and complete quantum generalization of the classical root-mean-square
error \cite{Oza19}.

We compare their research with ours as follows.
\begin{itemize}
\item[(i)] 
Their setup and approximation are the same as ours and they used the same Hamiltonian as in our research.
\item[(ii)] 
In both papers, the orbital state of the particle is assumed to be the pure state 
where the mean values of its position and momentum are zero.
We assume that the correlation term of a Gaussian pure state is not necessarily zero, whereas 
they assumed that the orbital state is a Gaussian pure state with no correlation.
\item[(iii)]
We evaluate the tradeoff between the error and disturbance, whereas they compared the error with the 
inner product $I$ of the emerging wave 
functions expressing formal distinguishability.
In addition, we obtain the range of error and disturbance under the condition 
that the orbital state is a 
Gaussian pure state whose correlation term is 
not necessarily zero.
\end{itemize}

\section{Conclusion} \label{conclusion}
Stern-Gerlach measurements, originally performed by Gerlach and Stern  \cite{SG22a,SG22b,SG22c},
have been discussed for a long time as a typical model
or a paradigm of quantum measurement \cite{Boh51}.
As Heisenberg's uncertainty principle suggests,  
Stern-Gerlach measurements of one spin component inevitably disturb its orthogonal component,
and Heisenberg's EDR \eq{HEDR} has been commonly believed to be its precise quantitative expression.
However, general quantitative relations between error and disturbance in arbitrary quantum 
measurements have been extensively investigated over the past two decades
and universally valid EDRs have been obtained to reform
Heisenberg's original EDR
(see, e.g., \cite{Oza03a,ESSGO12,Bra13,BLW14,Oza19} and references therein).

Here we investigated the EDR for this familiar 
class of measurements in light of the general theory 
leading to the universally valid EDR relations.
We have determined the range of possible values of the error and disturbance 
achievable by arbitrary Stern-Gerlach apparatuses, assuming that the orbital state 
is a Gaussian state.
Our result is  depicted in Fig.~\ref{figed} and the boundary of the
error-disturbance region is given in  \Eq{edSG} as a closed formula.
The result shows that the error-disturbance region of Stern-Gerlach measurements
occupies a near-optimal subregion of the universally valid error-disturbance 
region for arbitrary measurements.
It can be seen that one of the earliest methods of quantum 
measurement violates Heisenberg's EDR \eq{HEDR} in a broad 
range of experimental parameters.
Furthermore, we found a class of initial orbital states 
in which the error can be minimized an arbitrarily small amount by the screen 
at a finite distance from the magnet in contrast to the conventional 
assumption that the error decreases asymptotically.

The relation for the general class of states beyond Gaussian states 
is left to future study.
In addition, we also leave it to future research to analyze 
more realistic models, for example, a model described by the magnetic field 
satisfying Maxwell's equations \cite{BC00,PBC05} 
or a model considering the decoherence of the particle during 
the measuring process \cite{Dev15}.

Our results will contribute to answer the question as to
how various experimental parameters can be controlled to 
achieve the ultimate limit. 
We expect that the present study will provoke further experimental 
studies.

\acknowledgments
The authors thank Kazuya Okamura for helpful discussions.
This work was partially supported by JSPS KAKENHI, Grants No.~JP26247016,  
No.~JP17K19970, and the IRI-NU collaboration.

\appendix

\section{Error and disturbance in quantum measurements
 \label{gted}}
 In this appendix, we review the general theory of error and disturbance in
quantum measurements developed in \cite{Oza04,Oza19}.

\subsection{Classical root-mean-square error}

Let us consider the classical case first.
Recall the 
root-mean-square (rms) error introduced by Gauss \cite{Gau21}. 
Consider a measurement of the value $x$ of a quantity $X$ 
by actually observing the value $y$ of a meter quantity $Y$.
Then the
error of this measurement is given by
$y-x$.
If these quantities obey a joint probability distribution $\mu(x, y) $, 
then the rms error $\varepsilon _{G}(\mu )$ is defined as
\begin{equation}
\varepsilon _{G}(\mu )
=
\left(
 \sum_{x,y} (y-x)^2\, \mu  (x, y)
\right)^{1/2}.
\end{equation} 

\subsection{Quantum measuring processes}
We consider a quantum system $\bS$ described by a finite-dimensional 
Hilbert space $\cH$.
We assume that every measuring apparatus for the system $\bS$ 
has its own output variable $\bx$.
The statistical properties of the apparatus $\bA(\bx)$ having the output 
variable $\bx$ are determined by
(i) the probability distribution $\Pr\{\bx=m\|\rho\}$ of $\bx$ for the
input state $\rho$, and (ii) the output state $\rho_{\{\bx=m\}}$ given
the outcome $\bx=m$.

A measuring process of the apparatus $\bA(\bx)$ measuring  $\bS$
is specified by a quadruple $\bM=(\cK,\kxi,U,M)$ consisting of a Hilbert space $\cK$ 
describing the probe system $\bP$, 
a state vector $\kxi$ in $\cK$ describing the initial state of $\bP$, 
a unitary operator $U$ on $\cH\otimes\cK$ describing the time
evolution of the composite system $\bS+\bP$ during the measuring interaction,
and an observable, $M$, called the meter observable, 
of $\bP$ describing the meter of the apparatus.

The instrument of the  measuring process $\bM$ is defined as
a completely positive map valued function $\cI$ given by
\beq
 \cI(m)\rho=\Tr_{\cK}[(\1\otimes P^{M}(m))U(\rho\otimes\ketbra{\xi})U^{\dagger}]
\eeq
for any state $\rho$ and real number $m$.
The statistical properties of the apparatus $\bA(\bx)$ are determined by the
instrument $\cI$ of $\bM$  as
\begin{align}
\Pr\{\bx=m\|\rho\}&=
\Tr[\cI(m)\rho],\\
\rho_{\{\bx=m\}}&=
\frac{\cI(m)\rho}{\Tr[\cI(m)\rho]}.
\end{align}

The non-selective operation $T$ of $\bM$ is defined by
\deq{
T=\sum_{m\in\R}\cI(m).
}
Then we have
\deq{
T(\rho)=\Tr_{\cK}[U(\rho\otimes\ketbra{\xi})U^{\dagger}].
}

See Refs.~\cite{Oza84,Oza89,Oza04} for detailed descriptions of measuring processes
and instruments.

\subsection{Heisenberg picture}
In the measuring process $\bM$, we suppose that the measuring interaction is turned on
from time $t=0$ to time $t=\t$.
Then, the outcome $\bx=m$ of the apparatus $\bA(\bx)$ described by
the measuring process $\bM$ is defined as the outcome $m$ of the meter 
measurement at time $t=\t$.
To describe the time evolution of the composite system $\mathbf{S}+\mathbf{P}$ 
in the Heisenberg picture, let
\beqa
\begin{array}{rclcrclc}
A(0)&=&A\otimes \1, &\quad&  A(\t)&=&U^{\dagger}A(0)U,\\
B(0)&=&B\otimes \1,  &\quad&  B(\t)&=&U^{\dagger}B(0)U,  \\
M(0)&=&\1\otimes M, &\quad& M(t_0)&=&U^{\dagger}M(0)U,
\end{array}
\eeqa
where $A$ and $B$ are observables of $\bS$.

Then, the POVM $\PM $ of $\mathbf{M}$ is defined as
\begin{equation}
\PM(m)
=
\av{\xi|P^{M(\t)}(m)|\xi}
\end{equation}
and satisfies
\begin{equation}
\Pr\{\bx=m\|\rho\}=\Tr[\PM(m)\rho].
\end{equation}

The $n$-th moment operator of $\PM$ for $n=1,\ldots,n$ is defined by
\beq
\hPM^{(n)}
=\left\langle \xi \left|  M(t_0 )^n\right| \xi \right\rangle. 
\eeq
The dual non-selective operation $\TM$ of $\bM$ is defined by
\beq
\TM(B)=\av{\xi|B(\t)|\xi}
\eeq
for any observable $B$ of $\bS$ and satisfies
\deq{
\Tr\left\{\left[T^{*}(B)\right]\rho\right\}=\Tr\left\{B\left[T(\rho)\right]\right\}
}
for any observable $B$ and state $\rho$.

\subsection{Measurement of observables}
If the observables $A(0)$ and $M(t_0)$ commute in the initial state 
$\RS \otimes | \xi \rangle \langle \xi | $, that is,
\begin{equation}
[ P^{A(0)}(a), P^{M(t_0)}(m)]
 (\RS \otimes \left|  \xi \rangle \langle \xi \right| )
=0
\end{equation}
for all $a,m\in\R$,
then their joint probability distribution 
$\mu^{\AM} (a, m)$ 
is defined as 
\begin{equation}
\mu^{\AM}(a, m)
=
\mathrm{Tr}
[
 P^{A(0)}(a) P^{M(t_0)}(m)\, (\RSX)
]
\end{equation}
 and satisfies 
\begin{equation}
\mathrm{Tr}
[f(A(0),M(t_0))(\RSX)]
=
\sum_{a,m}f(a,m)\,\mu(a,m)
\end{equation}
for any polynomial $f(A(0),M(t_0))$ of  $A(0)$ and $M(t_0)$.

We say that the measuring process $\mathbf{M}$ {\em accurately measures}
the observable $A$ in a state $\RS$ if $A(0)$ and $M(t_0)$ are
perfectly correlated in the state $ \RS \otimes \left|  \xi \rangle \langle \xi \right| $ \cite{05PCN,06QPC,Oza19}, 
namely, one of the following two equivalent 
conditions holds:
(i)  $A(0)$ and $M(t_0)$ commute in $\RS \otimes | \xi \rangle \langle \xi | $
and their joint probability distribution $\mu^{\AM}$ satisfies 
\begin{equation}
\sum_{a,m:a=m}\mu^{\AM}(a,m)=1
\end{equation}
 or (ii) for any $a,m\in\R$ with $a\ne m$,
 \begin{equation}
 \mathrm{Tr}\left[ \PM(m)P^{A}(a) \,\RS\right]=0.
 \end{equation} 
 
 Note that  $\nu(a,m):=\mathrm{Tr}\left[ \PM(m)P^{A}(a) \,\RS\right]$, called the
 weak joint distribution of $A(0)$ and $M(\t)$, always exists and
 is operationally accessible by weak measurement and post-selection \cite{Joz07,LW10}, but possibly
 takes negative or complex values.
 Since $\nu(a,m)$ is operationally accessible, our definition of accurate measurements is
 operationally accessible.
 
 \subsection{Quantum root-mean-square error}
 The noise operator $N(A, \mathbf{M})$
of the measuring process $\mathbf{M}$ for measuring $A$ 
is defined as
\begin{equation}
N(A, \mathbf{M})
=M(t_0)-A(0).
\end{equation}  
The (noise-operator based) quantum rms error 
$\epn(A, \mathbf{M},\rho )$ for measuring $A$ in $\rho$ by $\bM$
is defined as the root mean square of the 
noise operator, i.e., 
\begin{equation}
\epn(A, \mathbf{M},\rho )
= \left\{\mathrm{Tr}\left[N(A,\mathbf{M})^2 (\RSX)\right]\right\}^{1/2} .
\end{equation}

To argue the reliability of the error measure $\epn$ defined above, we consider the following 
requirements for any reliable error measures $\ep$ generalizing the classical root-mean-square 
error $\epg$ to quantify the mean error $\ep(A,\bM,\rho)$ of the measurement of an observable 
$A$ in a state $\rho$ described by a measuring process $\bM$ \cite{Oza19}.
\begin{itemize}
\item[\textit{(i)}]\textit{Operational definability.}
The error measure $\ep$ should be definable by the POVM $\PM$ of the measuring process $\mathbf{M}$
with the observable $A$ to be measured and the initial state $\RS$ of the measured system $\mathbf{S}$.
\item[\textit{(ii)}]\textit{Correspondence principle.}
In the case where $A(0)$ and $M(t_0)$ commute in $\RS\otimes\ketbra{\xi}$,
the relation
\begin{equation}
\varepsilon (A, \mathbf{M},\RS) = \varepsilon _{G} (\mu )
\end{equation}
holds for the joint probability distribution 
$\mu$ of $A(0)$ and $M(t_0 )$ in $\RS\otimes\ketbra{\xi}$.
\item[\textit{(iii)}]\textit{Soundness.}
If $\mathbf{M}$ accurately measures  $A$ in $\RS$, then 
$\varepsilon $ vanishes, i.e., 
$\varepsilon(A,\mathbf{M},\RS) =0$.
\item[\textit{(iv)}]\textit{Completeness.} 
If  $\varepsilon $ vanishes, then $\mathbf{M}$ accurately measures $A$ in $\RS$.
\end{itemize}

It was shown in \cite{Oza19} that the noise-operator-based quantum rms error $\ep=\epn$ 
satisfies requirements (i)--(iii),
so it is a sound generalization of the classical rms error.
However, as pointed out by Busch \textit{et al.}~\cite{BHL04}, 
$\ep=\epn$ may not satisfy the completeness requirement (iv) in general.
To improve this point, in Ref.~\cite{Oza19} a modification 
of the noise-operator-based quantum rms error $\epn$
was introduced to satisfy all the requirements (i)--(iv) as follows.  
The locally uniform quantum rms error $\epu$ is defined by
\deq{
\epu(A,\bM,\rho)=\sup_{t\in\R}\epn(A,\bM,e^{-itA}\rho e^{itA}).
}
Then $\ep=\epu$ satisfies all the requirements (i)--(iv) including completeness.
In addition to (i)--(iv), the new error measure  $\epu$ has the following two properties.
\begin{itemize}
\item[\textit{(v)}] {\em Dominating property.}  The error measure $\epu$ dominates $\epn$, i.e., 
$\epn(A,\bM,\rho)\le\epu(A,\bM,\rho)$.
\item[\textit{(vi)}] {\em Conservation property for dichotomic measurements.}
The error measure $\epu$ coincides with $\epn$ for dichotomic measurements,
i.e., $\epu(A,\bM,\rho)=\epn(A,\bM,\rho)$ if $A(0)^2=M(\t)^2=\1$.
\end{itemize}

By property (v) the new error measure $\epu$ maintains the previously obtained universally
valid EDRs \cite{Oza03a,Bra13,Oza14}. 
In this paper we consider the measurement of a spin component $\si_z$ of a spin-$1/2$ 
particle using a dichotomic meter observable $M$, i.e., $M^2=\1$, so by property
(vi) of $\epu$ we conclude that the noise-operator-based quantum rms error $\epn$ satisfies all
the requirements (i)--(iv) for our measurements under consideration without modifying it to be $\epu$.

As shown in \Eq{noise-operator}, in our model of the Stern-Gerlach measurement,  the Heisenberg 
observables $A(0)$ and $M(t_0)$ commute,
so the error measure satisfying (i) and (ii) is uniquely determined as the
(noise-operator-based) quantum rms error.

Busch \textit{et al.}~\cite{BLW14} criticized the use of the noise-operator-based
quantum rms error, by comparing it with the error measure based on the Wasserstein 
2-distance,  another error measure defined as the Wasserstein 2-distance between the
probability distributions of $A(0)$ and $M(t_0)$.
As shown in Ref.~\cite{Oza19}, the error measure based on the
Wasserstein 2-distance or based on any distance between the
probability distributions of $A(0)$ and $M(t_0)$
satisfies (i) and (iii) but does not satisfy (ii) or (iv), 
so the discrepancies between those two measures do not lead to the conclusion 
that the noise-operator-based quantum rms error is less reliable than the error measured based 
on the Wasserstein 2-distance or based on any distance between probability distributions
of $A(0)$ and $M(t_0)$.

In what follows, where no confusion may occur, we will write $\ep(A)=\ep_{NO}(A)$ for brevity.

\subsection{Disturbance of observables}
We say that the measuring process $\mathbf{M}$ 
does not disturb the observable $B$ in a state $\RS$ if $B(0)$ and $B(\t)$ are
perfectly correlated in the state $\RSX$ \cite{05PCN,06QPC,06NDQ},
namely, one of the following two equivalent 
conditions holds: 
(i) $B(0)$ and $B(\t)$ commute in $\RSX$
and their joint probability distribution $\mu^{\AM}$ satisfies 
\begin{equation}
\sum_{b,b':b=b'}\mu^{\AM}(b,b')=1
\end{equation}
or (ii) for any $b,b'\in\R$ with $b\ne b'$,
 \begin{equation}\label{eq:WJD}
 \mathrm{Tr}\left[ P^{B(\t)}(b')P^{B(0)}(b)\RSX\right]=0.
 \end{equation} 
 
Note that the left-hand side of \Eq{WJD} is 
called the weak joint distribution of $B(0)$ and $B(\t)$ and always exists,
possibly taking negative or complex values.
The weak joint distribution is operationally accessible by weak measurement of $B(0)$ 
and post selection for $B(\t)$ \cite{Joz07,LW10}.
 Thus, our definition of non disturbing measurement is operationally accessible.
  
\subsection{Quantum root-mean-square disturbance}

For any observable $B$ of the system $\bS$,
the disturbance operator $D(B, \mathbf{M})$ 
for the measuring process $\mathbf{M}$
causing the observable $B$ is defined as the change of the observable $B$ 
during the measurement, i.e., 
\begin{equation}
D (B,\mathbf{M}) = B(t_0) - B(0).
\end{equation}
Similarly to the quantum rms error, the 
quantum rms disturbance
$\eta(B, \mathbf{M},\rho )$ of $B$ in $\rho$ caused by $\bM$ is defined as the rms  of the disturbance operator, i.e., 
\begin{equation}
\eta(B, \mathbf{M} )
=
 \left\{
\mathrm{Tr}
[D(B,\mathbf{M})^2 (\RSX)]
\right\}^{1/2} .
\end{equation}
The quantum rms disturbance  $\eta$ has properties analogous to the (noise-operator-based) quantum rms 
error as follows.  
\begin{itemize}
\item[\textit{(i)}]
\textit{Operational definability.}
The quantum rms disturbance $\eta$ is definable by the non selective operation $T$
of the measuring process $\mathbf{M}$,
the observable B to be disturbed, and the initial state $\RS$ of the measured system $\mathbf{S}$.
\item[\textit{(ii)}]
\textit{Correspondence principle.}
In the case where $B(0)$ and $B(t_0)$ commute in $\RS\otimes\ketbra{\xi}$,
the relation
\begin{equation}
\eta(B, \mathbf{M},\RS) = \varepsilon _{G} (\mu )
\end{equation}
holds for the joint probability distribution 
$\mu$ of $B(0)$ and $B(t_0 )$ in $\RS\otimes\ketbra{\xi}$.
\item[\textit{(iii)}]
\textit{Soundness.} If $\mathbf{M}$ does not disturb $B$ in $\RS$, then $\eta $ vanishes.
\item[\textit{(iv)}]
\textit{Completeness for dichotomic observables.}
In the case where  $B^2=\1$, 
if  $\eta $ vanishes, then  
$\mathbf{M}$ does not disturb $B$ in $\RS$.
\end{itemize}

Korzekwa \textit{et al.}~\cite{KJR14} criticized the use of the 
operator-based quantum rms disturbance relying on their definition
of non disturbing measurements.  They define non disturbing measurements
in a system state $\rho$ as measurements satisfying that $B(0)$ and $B(t_0)$ 
have identical probability distributions for the initial state $\rho\otimes\ketbra{\xi}$.
They claimed that the operator-based quantum rms disturbance does
not satisfy the soundness requirement based on their definition of non disturbing 
measurements.  However, the conflict can be easily reconciled, since their definition of 
non disturbing measurement is not strong enough, i.e., they call a measurement 
non disturbing even when the disturbance is operationally detectable.  In fact, they 
supposed that the projective measurement of $A=\si_z$ of a spin-1/2 particle 
in the state $\ket{\si_z=+1}$ does not disturb the observable $B=\si_x$.  However,
this measurement really disturbs the observable $B=\si_x$.  In fact, we have 
\deqs{
\lefteqn{\av{\psi,\xi|P^{B(\t)}(b')P^{B(0)}(b)|\psi,\xi}}\quad\\
&=|\av{\si_z=+1|\si_x=b'}|^2|\av{\si_z=+1|\si_x=b}|^2.
}
Thus, $B(0)$ and $B(\t)$ have the same probability distribution, i.e., 
\deq{
\av{\psi,\xi|P^{B(\t)}(b)|\psi,\xi}=\av{\psi,\xi|P^{B(0)}(b)|\psi,\xi},
\label{eq:identical_probability}
}
but the weak joint distribution operationally detects the disturbance on $B$, i.e., 
\deq{
\av{\psi,\xi|P^{B(\t)}(-1)P^{B(0)}(+1)|\psi,\xi}=1/4.
\label{eq:disturbing}
}
In this case, we have $\et(B,\bM,\rho)=\sqrt{2}\not=0$ (see \cite{05UUP}p.~S680).  However, this does not mean that $\et$
does not satisfy the soundness requirement, since $\bM$ disturbs $B$ in $\rho$ according to
\Eq{disturbing}. The detail will be discussed elsewhere.

\subsection{Universally valid error--disturbance relations}

In the following,  where no confusion may occur, we abbreviate  $\varepsilon(A, \mathbf{M},\RS) $ 
as  $\varepsilon(A)$  and 
$\eta(B,\mathbf{M},\RS)  $ as $\eta(B) $.

In Ref.~\cite{Oza03a} Ozawa derived the relation 
\begin{multline}
\varepsilon (A) \eta (B) \! + \! \varepsilon  (A) \sigma (B) 
\! + \! \sigma (A) \eta (B)
\geq 
\frac{1}{2}|\mathrm{Tr}([A, B]\RS)|,
\end{multline}
holding for any pair of observables $A$ and $B$, state $\ket{\psi}$, 
and measuring process $\mathbf{M}$. 
Subsequently, Brancirard \cite{Bra13} and Ozawa \cite{Oza14} obtained a stronger EDR given by
\begin{eqnarray}
\lefteqn{\varepsilon(A)^2 \sigma(B)^2  + \sigma(A)^2 \eta (B)^2 \Bigr.} \notag \\
& & \Bigl. + 2  \varepsilon (A) \eta (B) \sqrt{ \sigma(A)^2\sigma(B)^2-D_{AB}^2}
\geq D_{AB}^2,
\quad
\label{eq:BEDR}
\end{eqnarray}
where 
\begin{equation}
D_{AB}=\frac{1}{2}\mathrm{Tr}(\left| \sqrt{\RS }[A, B] \sqrt{\RS }\right|).
\end{equation}
In the case where
$A^2=B^2=\1$ and $M^2 = \1$, 
relation (\ref{eq:BEDR}) can be strengthened as \cite{Bra13,Oza14}
\begin{equation}
\hat \varepsilon (A)^2  + \hat \eta (B)^2+ 2  \hat \varepsilon(A) \hat \eta(B) \sqrt{ 1-D_{AB}^2}
\geq D_{AB}^2,
\label{eq:BEDRM}
\end{equation}
where
$\hat \varepsilon (A)=\epsilon  (A)\sqrt{1-\frac{\epsilon  (A)^2}{4}}$
and
$\hat \eta          (B)=\eta      (B)\sqrt{1-\frac{\eta      (B)^2}{4}}$.
In the case where
\begin{equation}
A = \sigma _z, \quad B = \sigma _x, \quad 
\left\langle \sigma _z(0) \right\rangle _{\RS } = \left\langle \sigma _x(0) \right\rangle _{\RS } = 0, 
\label{eq:condition0-A}
\end{equation}
the inequality (\ref{eq:BEDRM}) is reduced to the tight relation \cite{Bra13,Oza14}
\begin{equation}\label{eq:BOEDRS-A}
\left[
 \varepsilon  (\sigma _z) ^2 -2
\right]^2
+
\left[
 \eta (\sigma _x) ^2 -2
\right]^2
\leq 4,
\end{equation}
as depicted in FIG \ref{borel}.

\section{Gaussian wave packets}\label{Gaussianapp}
In this appendix, we review the relations between Gaussian states and inequalities. 
Let $Z$ and $P$ be the canonical position and momentum observables, 
respectively, of a one-dimensional quantum system.
These observables satisfy the usual canonical commutation relation $[Z,P] = i \hbar $.
Here we consider only a vector state denoted by $\psi$.
However, some of the results in this appendix can easily be generalized to mixed states.

\subsection{Schr\"{o}dinger inequality}\label{se:schr_ineq}
For the variances of the position and momentum, the following inequality holds \cite{Sch30}:
\begin{equation}\label{schr_ineq}
\mathrm{Var} _{\psi }(Z)
\mathrm{Var} _{\psi }(P)
\geq
\frac{
\left(
 \langle
 \{Z, P\}
 \rangle _{\psi }
 -2
 \langle
 Z
 \rangle _{\psi}
 \langle
 P
 \rangle _{\psi}
\right)^2
+
\hbar ^2
}
{4}.
\end{equation}
Inequality (\ref{schr_ineq}) is known as the Schr\"{o}dinger inequality.
The proof proceeds as follows.
First, we consider the case $\langle Z \rangle _{\psi } = \langle P \rangle _{\psi } =0$.

Then we have
\begin{eqnarray}
\Im \langle Z \psi, P \psi \rangle
=
\frac{1}{2 i}
\langle \left[ Z, P \right]\rangle _{\psi }
=
\hbar /2, \\
\Re \langle Z \psi, P \psi \rangle
=
\frac{1}{2}
\langle \{ Z, P \} \rangle _{\psi }.
\end{eqnarray}
Consequently, we have
\begin{equation}
\vert \langle Z \psi , P \psi \rangle \vert ^2
=
\frac{
\left(
 \langle
 \{Z, P\}
 \rangle _{\psi }
\right) ^2
+
\hbar^2}{4}.
\end{equation}
On the other hand, according to the Cauchy-Schwarz inequality,
\begin{equation}
\vert \langle Z \psi, P \psi \rangle \vert ^2
\leq
\langle Z^2 \rangle _{\psi }
\langle P^2 \rangle _{\psi }
=
\mathrm{Var}_{\psi}(Z)
\mathrm{Var}_{\psi}(P).
\end{equation}
Hence, the Schr\"{o}dinger inequality (\ref{schr_ineq}) holds if
$\langle Z \rangle _{\psi } = \langle P \rangle _{\psi } =0$
holds.
We can obtain the proof for the general case by substituting $Z$ and $P$ into 
$Z - \langle Z \rangle _{\psi}$ and $P - \langle P \rangle _{\psi}$, respectively.
This concludes the proof.

The equation in this inequality holds if and only if
\begin{equation}
\left( Z -  \langle Z \rangle _{\psi } \right) \psi = c \left( P -  \langle P \rangle _{\psi } \right) \psi
\label{schroeqcondition}
\end{equation}
for some complex number $c$.
From the condition above, we obtain the differential equation for the wave function as
\begin{equation}
\frac{d}{dz}\psi(z)
=
-2k
\left[
 z
 -
 \left(
  \langle Z \rangle _{\psi }
  +
  \frac{i}{2 \hbar k}
  \langle P \rangle _{\psi }
 \right)
\right]
\psi (z),
\end{equation}
where $k$ is a complex number.
Therefore, we have
\begin{equation}
\psi (z)
=
A
\exp
\left(
 -k
 \left[
  z
  -
  \left(
   \langle Z \rangle _{\psi }
   +
   \frac{i}{2 \hbar k}
   \langle P \rangle _{\psi }
  \right)
 \right]^2
\right), \label{schrodingerwaveequation}
\end{equation}
where $A$ is a constant.
Since the wave function should be normalizable, the constant $k $ must satisfy $\Re k > 0$.

\subsection{Kennard inequality}
The inequality, 
which is known as the Kennard inequality \cite{Ken27}
\begin{equation}
\mathrm{Var}_{\psi}(Z)\mathrm{Var}_{\psi}(P) \geq \hbar ^2 /4, \label{RKineq}
\end{equation}
can be derived from the Schr\"{o}dinger inequality (\ref{schr_ineq}). The equality in Eq.~(\ref{RKineq}) holds if and only if 
$2i \hbar k \left( Z -  \langle Z \rangle _{\psi } \right) \psi
=
\left( P -  \langle P \rangle _{\psi } \right) \psi $
for some positive real number $k$.
A wave function $\psi$ satisfies the equality in the Kennard inequality (\ref{RKineq}) 
if and only if $\psi $ has the form
\begin{equation}
\psi (z)
=
A
\exp
\left(
 -k
 \left[
  z
  -
  \left(
   \langle Z \rangle _{\psi }
   +
   \frac{i}{2 \hbar k}
   \langle P \rangle _{\psi }
  \right)
 \right]^2
\right) \label{Kennardwaveequation}
\end{equation}
for some positive real number $k$.
This wave function has the same form as that of 
Eq.~(\ref{schrodingerwaveequation}) except for the condition of the constant $k$,
i.e., the constant $k$ in Eq.~(\ref{schrodingerwaveequation}) is a complex number with a positive real part
whereas the constant $k$ in Eq.~(\ref{Kennardwaveequation}) is a positive real number.
The state in Eq.~(\ref{Kennardwaveequation}) is known as the minimum-uncertainty state.

\subsection{Squeezed state}
For any two complex numbers $\mu $ and $\nu$ satisfying $\vert \mu \vert ^2 - \vert \nu \vert ^2 =1$,
the squeezed operator $c _{\mu, \nu}$ is defined as 
\begin{equation}
c _{\mu, \nu} := \mu a + \nu a^{\dag},
\end{equation}
where $a$ and $a^{\dag }$ are the annihilation and creation operators, respectively.
\begin{equation}
a := \sqrt{\frac{m \omega }{2\hbar }}Z + i \sqrt{\frac{1 }{2\hbar m \omega }} P. \label{annihilation}
\end{equation}
Here, $m $ and $\omega $ are the mass and angular frequency of the corresponding
harmonic oscillator, respectively.
A coherent state \cite{Gla63} is defined as the eigenstate of the annihilation operator $a$ in Eq.~(\ref{annihilation}).
A squeezed state \cite{Yue76} is defined as the eigenstate of squeezed operator $c_{\mu, \nu}$,
\begin{equation}
c _{\mu , \nu } \psi = \lambda \psi.
\end{equation}
By this definition, the wave function of every squeezed state satisfies the differential equation
\begin{equation}
\left[
 \left(
  \mu + \nu
 \right)
 \sqrt{\frac{m \omega }{2 \hbar}}
 z
 +
 \left(
  \mu - \nu
 \right)
 \sqrt{\frac{\hbar}{2 m \omega}}
 \frac{d}{dz}
\right]
\psi (z)
=
\lambda \psi (z).
\end{equation}
The solution of this differential equation is
\begin{equation}
\psi (z)
:=
A
\exp
\left[
 -\frac{m \omega }{2 \hbar }\frac{\mu +\nu}{\mu -\nu}
 \left(
  z -
  \sqrt{\frac{2 \hbar }{m \omega }}
  \frac{\lambda }{\mu -\nu }
 \right)^2
\right]. \label{sqeezedwavefunction}
\end{equation}
Hence, the equality in the Schr\"{o}dinger inequality (\ref{schr_ineq}) holds for squeezed states.

Next let us consider the relation between these parameters and the mean values of the position 
and momentum.
By comparing the two formulas, (\ref{schrodingerwaveequation}) and (\ref{sqeezedwavefunction}), 
we have
\begin{equation}
\langle Z \rangle _{\psi }
+
\frac{i}{m \omega }\frac{\mu - \nu }{\mu + \nu}
\langle P \rangle _{\psi }
=
\sqrt{\frac{2 \hbar }{m\omega }}
\frac{\lambda }{\mu-\nu}.
\end{equation}
Taking the imaginary part, we have
\begin{eqnarray}
\langle P \rangle _{\psi }
&=&
\sqrt{2 \hbar m \omega } \vert \mu + \nu \vert ^2 \Im \left( \frac{\lambda }{\mu - \nu } \right), \\
\langle Z \rangle _{\psi }
&=&
\sqrt{\frac{2 \hbar }{m\omega }}
\Re \left( \frac{(\mu + \nu )(\mu ^* - \nu^*)}{\mu- \nu } \lambda \right).
\end{eqnarray}
Next, let us calculate the variances of the position and momentum and the correlation 
$\langle \{ Z , P \} \rangle _{\psi }$.
Setting $\tilde{z}=z - \langle Z \rangle _{\psi }$, we have
\begin{widetext}
\begin{equation}
\mathrm{Var}(Z) 
=
|A|^2\!
\int _{-\infty }^{\infty  }\!
\tilde{z}^2
\exp
\left\{
 -\frac{m \omega }{\hbar }
 \Re\!\!
 \left[
  \frac{\mu- \nu }{\mu +\nu }
  \left(
   \frac{\mu+ \nu }{\mu -\nu }
   \tilde{z} + \frac{i}{m \omega }\langle P \rangle _{\psi }
  \right)^2
 \right]
\right \}
d \tilde{z}
=
\frac{\hbar }{2m \omega } \vert \mu - \nu \vert ^2 .
\end{equation}
To calculate the variance of the momentum, it is convenient to obtain the Fourier transform of 
the wave function 
$\tilde{\psi } (\tilde{z}) := \psi (\tilde{z}+ \langle Z \rangle _{\psi })$,
\begin{equation}
\hat{\psi }(p) 
=
\frac{1}{\sqrt{2 \pi \hbar }}
\int _{-\infty }^{\infty }
\tilde{\psi } (\tilde{z}) \exp(i p \tilde{z} / \hbar ) d \tilde{z} 
=
\hat{A} \exp 
\left[
 - \frac{1}{2 \hbar m \omega}
 \frac{\mu - \nu }{\mu + \nu }
 \left(
  p - \langle P \rangle _{\psi }
 \right)^2
\right],
\end{equation}
where $\hat{A}$ is the normalization constant.
Consequently, we have 
\begin{equation}
\mathrm{Var}(P) 
=
\langle (P - \langle P \rangle _{\psi })^2 \rangle _{\psi}
=
\vert \hat{A} \vert ^2
\int _{-\infty } ^{\infty}
\tilde{p}^2
\exp
\left[
 -\frac{1}{\hbar m \omega }
 \Re
 \left(
  \frac{\mu - \nu }{\mu + \nu }
 \right)
 \tilde{p}^2
\right]  d\tilde{p}
=
\frac{\hbar m \omega }{2}
\vert \mu + \nu \vert ^2.
\end{equation}
Finally, we calculate the correlation term
\begin{align}
\lefteqn{
\langle \{Z - \langle Z \rangle _{\psi }, P - \langle P \rangle _{\psi } \} \rangle _{\psi}}\quad \notag\\
&=
\langle \{Z - \langle Z \rangle _{\psi }, P \} \rangle _{\psi} 
=
2 \Re \langle \tilde{Z} \psi, P \psi \rangle 
=
2 \Re
\left\{ \vert A \vert ^2  i m \omega 
 \int _{-\infty }^{\infty }
 \frac{\mu + \nu }{\mu - \nu }
 \tilde{z}^2 \!
 \exp \!
 \left[
  -\frac{m \omega }{\hbar }
  \Re
  \left(
   \frac{\mu + \nu }{\mu -\nu } \tilde{z}^2
  \right)
 \right]
 d \tilde{z}
\right\}
=
2 \hbar \Im(\mu ^* \nu ).
\end{align}
The coherent state is defined as the eigenstate of the annihilation operator. 
Using the results of the calculation above, the corresponding wave function is
\begin{equation}
\psi (z)
=
A
\exp
\left[
 -\frac{m \omega }{2 \hbar }
 \left(
  z -
  \sqrt{\frac{2 \hbar }{m \omega }}
  \lambda
 \right)^2
\right],
\end{equation}
where $\lambda $ is the corresponding eigenvalue of the annihilation operator.
Thus, every coherent state satisfies the equation in the Schr\"{o}dinger inequality (\ref{schr_ineq}) and 
the Kennard inequality (\ref{RKineq}).

\end{widetext}
Since $\dfrac{\mu + \nu }{\mu - \nu }$ moves all over the right half plane of the complex plane
as $\mu $ and $\nu $ move all over the complex plane satisfying $\vert \mu \vert ^2 - \vert \mu \vert ^2 =1$,
the union of all squeezed states and coherent states coincides with the states that satisfy 
the Schr\"{o}dinger inequality  (\ref{schr_ineq}), namely, $\mathcal{G}$.

\subsection{Contractive state}
\label{se:contractive}
The contractive state was introduced by Yuen \cite{Yue83}
as a squeezed state whose correlation term is negative. 
This state contracts during some period of time if it evolves freely. To see this, 
let us calculate the variance of the position in the Heisenberg picture.
The position operator $Z(t)$ at time $t$ in the Heisenberg picture is
\begin{align}
Z(t)
&=
\exp \left[-\frac{t}{2 i \hbar m } P(t)^2 \right]Z(0) \exp \left[\frac{t}{2 i \hbar m } P(t)^2 \right] 
\notag \\
&=
Z(0) + \frac{t}{m} P(0).
\end{align}
Hence, we have
\begin{align}
& \mathrm{Var} _{\psi }\left[ Z(t) \right] \notag \\
&=
\left\langle 
\left(
 Z(0) + \frac{t}{m} P(0)
 -
 \langle
 Z(0) + \frac{t}{m} P(0)
 \rangle _{\psi }
\right)^2
\right\rangle _{\psi } \notag \\
&=
\frac{t^2}{m^2}
\mathrm{Var} _{\psi } \left[ P(0) \right]
+
\mathrm{Var} _{\psi } \left[ Z(0) \right] \notag \\
&+
\frac{t}{m}
\left\langle
\{ Z(0) - \langle Z(0) \rangle _{\psi } , P(0) - \langle P(0) \rangle _{\psi } \} 
\right\rangle _{\psi }.
\end{align}
Therefore, if the state is a contractive state, the variance of the position contracts until the time 
\begin{equation}
t=
-
\frac
{
 m
 \langle
  \{
   Z(0)
   -
   \langle Z(0)\rangle _{\psi } ,
   P(0)
   -
   \langle P(0) \rangle _{\psi }
  \}
 \rangle _{\psi }}
 {2 \langle P(0)^2 \rangle _{\psi } }.
\end{equation}

\subsection{Covariance matrix formalism}
Recently, the covariance matrix was used to characterize Gaussian states \cite{WPGCRSL12}.
For a single-mode Gaussian state,
\begin{equation}
\psi (z)
=
A
\exp
\left(
 -k
 \left[
  z
  -
  \left(
   \langle Z \rangle _{\psi }
   +
   \frac{i}{2 \hbar k}
   \langle P \rangle _{\psi }
  \right)
 \right]^2
\right),
\end{equation}
the covariance matrix $V$ is defined as
\begin{align}
	V
	=&
	\left(
		\begin{array}{c c}
			\mathrm{Var}_{\psi } \left( Z \right) & \mathrm{Cor}_{\psi} (Z, P) \\
			\mathrm{Cor}_{\psi} (Z, P)  & \mathrm{Var}_{\psi } \left( P \right)
		\end{array}
	\right) \notag \\
	=&
	\left(
		\begin{array}{c c}
			\left[4 \mathrm{Re}(k) \right]^{-1}                  & -\frac{\hbar \mathrm{Im}(k)}{\mathrm{Re}(k)} \\
			-\frac{\hbar \mathrm{Im}(k)}{\mathrm{Re}(k)} & \frac{\hbar ^2 \left|k \right|^2}{\mathrm{Re}(k)}
		\end{array}
	\right).
\end{align}
Here, we used the abbreviation,
\begin{equation}
\mathrm{Cor}_{\psi} (Z, P) = \langle \{Z - \langle Z \rangle _{\psi }, P - \langle P \rangle _{\psi } \} \rangle_{\psi}.
\end{equation}

\subsection{Summary}
\begin{figure}[htb]
\begin{center}
\includegraphics[width=0.5\textwidth, viewport=20 360 770 690]{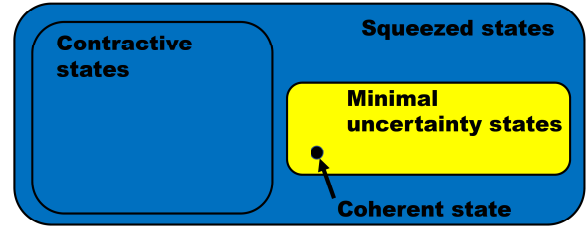}
\caption{
  Inclusion relation of the subsets of wave functions. A wave function is in the yellow region if and only if the equality 
  in the Kennard inequality holds. 
  A wave function is in the blue or yellow region if and only if the equality in the Schr\"{o}dinger inequality holds.
}
\label{squeezedstatesfig}
\end{center}
\end{figure}

We have discussed the relation between the inequalities and the subclasses of Gaussian states 
whose wave functions are of the form
\begin{equation}
\psi (z)
=
A \exp
\left(
 -k
 \left[
  z - (\langle Z \rangle _{\psi } + \frac{i}{2 \hbar k} \langle P \rangle _{\psi })
  \right]^2
\right)
\end{equation}
and 
obtained the relations shown in Table. \ref{table:data_type}.
Figure~\ref{squeezedstatesfig} represents the inclusion relation between the subsets of the set of Gaussian 
wave packets.
\begin{table}[hbtp]
  \caption{Classification of Gaussian states in terms of the parameter $k$.}
  \label{table:data_type}
  \centering
  \begin{tabular}{ccc}
    \hline
    $k$
    & Type of state & 
    \begin{tabular}{c}
    Inequality whose \\
    equality holds
    \end{tabular}
    \\
    \hline \hline
    $\Re k > 0$
    &
    Squeezed
    &
    Schr\"{o}dinger
    \\ \\
    \begin{tabular}{c}
     $\Re k > 0$ and \\
     $\Im k > 0$ 
    \end{tabular}
    &
    Contractive
    &
    Schr\"{o}dinger
    \\ \\
    \begin{tabular}{c}
    $\Re k > 0$ and
    \\
    $\Im k = 0$ 
    \end{tabular}
    &
    Minimum uncertainty
    &
    Kennard
    \\ \\
    $k = \hbar$
    &
    Coherent
    &
    Kennard
    \\
    \hline
  \end{tabular}
\end{table}

\section{Time evolution of Gaussian wave packets}
In this appendix we discuss the time evolution of the probability density of a Gaussian wave 
packet during free evolution.
The wave function under consideration is the Gaussian wave packet derived in Appendix~\ref{Gaussianapp},
\begin{equation}
\psi (z)
:=
A \exp
\left(
 - k z^2
\right),
\end{equation}
where $k $ is a complex number with a positive real part. For simplicity, we consider only the case 
in which the mean values of the position and momentum are zero.
Applying the Fourier transform $\mathfrak{F}$ successively, we obtain
\begin{align}
&\exp
\left(
 \frac{t}{2 i \hbar m }
 P^2
\right)\psi (z) \notag \\
&=
\mathfrak{F} ^{-1}
\exp
\left(
 \frac{t }{2 i \hbar m }
 p ^2
\right)
\hat{A}
\exp
\left(
 - \frac{p^2}{4 k \hbar ^2}
\right) \notag \\
&=
\frac{\hat{A}}{\sqrt{2 \pi \hbar }}
\int _{-\infty } ^{\infty}
\exp
\left[
 \left(
  \frac{t}{2i \hbar m}
  -
  \frac{1}{ k \hbar ^2}
 \right)
 p^2
 -
 ipz/ \hbar
\right]
dp \notag \\
&=
N \exp
\left(
 -
 \frac{z^2}{k ^{-1} - \frac{2 \hbar t}{im}}
\right),
\end{align}
where $N$ is the normalization constant.
Thus, the probability density $\mathrm{Pr}(z)$ at time $t$ has the form
\begin{equation}
\mathrm{Pr}(z)
=
\vert N \vert ^2
\exp
\left(
 -r z^2
\right)
\end{equation}
for some positive real number $r$, that is, we have again obtained a Gaussian distribution.
Since the variance of the Gaussian distribution is
\begin{equation}
\left\langle
 Z(t) ^2 
\right\rangle _{\psi }
=
\left\langle
  \left[ Z(0) + \frac{t}{m}P(0) \right] ^2
\right\rangle _{\psi },
\end{equation}
we have
\begin{equation}
\mathrm{Pr}(z)
=
\vert N \vert ^2
\exp
\left(
 -\frac{z^2}{2 \langle \left( Z(0) + \frac{t}{m}P(0) \right) ^2 \rangle _{\psi }}
\right).
\end{equation}

\section{Relationship between the Heisenberg picture and the Schr\"{o}dinger picture}
Let us consider the relation between the Heisenberg picture and the Schr\"{o}dinger picture.
Consider the time evolution of quantum system $\bS$ described by 
$\mathcal{H}$.
Let $A$ be an observable of system $\bS$ and state $\psi$.
Denote by $E(A,\psi,t)$ the expectation value of the outcome of the measurement of 
observable $A$ at time $t$, provided system $\bS$ is in state $\psi$ at time 0.
In the Schr\"{o}dinger picture,  state $\psi(t)$ evolves in time $t$ 
as a solution of the Schr\"{o}dinger equation by the time evolution operator $U(t)$ 
as $\psi(t)=U(t)\psi$ with the initial condition $U(0)=\1$,
so $E(A,\psi,t)=\av{\psi(t),A\psi(t)}$ holds.
The unitary operator $U^{\mathrm{S}}(t_2, t_1) $ describing the time evolution 
from time $t = t_1 $ to $t = t_2 $ $(t_1 \leq t_2)$ in the Schr\"{o}dinger 
picture is defined by 
\beq
U^{\mathrm{S}}(t_2, t_1)=U(t_2)U^{\da}(t_1).
\eeq
Then we have 
\begin{align}
U^{\mathrm{S}}(t_2, t_1)\psi(t_1)&=\psi(t_2),\\
U^{\mathrm{S}}(t_3, t_2)U^{\mathrm{S}}(t_2, t_1)&=U^{\mathrm{S}}(t_3, t_1).
\end{align}

In the Heisenberg picture, observable $A(t)$ evolves in time $t$ 
by the time evolution operator $U(t)$ as $A(t)=U(t)^{\da}AU(t)$,
so $E(A,\psi,t)=\av{\psi,A(t)\psi}$ holds.
The unitary operator $U^{\mathrm{H}}(t_2, t_1) $ describing the time evolution 
from time $t = t_1 $ to $t = t_2 $ $(t_1 \leq t_2)$ in the Heisenberg  
picture is defined by
\beq
U^{\mathrm{H}}(t_2, t_1)=U^{\da}(t_1)U(t_2).
\eeq
Then we have 
\begin{align}
U^{\mathrm{H}}(t_2, t_1)^{\da}A(t_1)U^{\mathrm{H}}(t_2, t_1)&=A(t_2),\\
\al^{\mathrm{H}}(t_3, t_2)\al^{\mathrm{H}}(t_2, t_1)&=\al^{\mathrm{H}}(t_3, t_1),
\end{align}
where 
\deq{
\al^{\mathrm{H}}(t_2, t_1)A=
U^{\mathrm{H}}(t_2, t_1)^{\da}AU^{\mathrm{H}}(t_2, t_1).
}

We have the following relations between the Schr\"{o}dinger picture and the Heisenberg picture:
\begin{eqnarray}
&U(t)=U^{\mathrm{S}}(t, 0 )
=
U^{\mathrm{H}}(t, 0 ). \label{0t}&\\
& U^{\mathrm{H}}(t_2, t_1 )
=
U(t_1)^{\da}U^{\mathrm{S}}(t_2, t_1 ) U(t_1).&
\end{eqnarray}
Let $f(A_1,\ldots,A_n,t,s)$ be a function of observables $A_1,\ldots,A_n$ and real numbers $t$ and $s$.
If
\deq{U^{\mathrm{S}}(t_2,t_1)=f(A_1,\ldots,A_n,t_1,t_2),}
then
\beq
 U^{\mathrm{H}}(t_2, t_1 )=f(A_1(t_1),\ldots,A_n(t_1),t_1,t_2).
\eeq
%}

\section{Solutions of Heisenberg equations of motion for 
$Z(t)$, $P(t)$, $\si_x(t)$, $\si_y(t)$, and $\si_z(t)$}
\label{se:App_SHEM}

To consider the time evolution from time $t= \Delta t$ to time 
$\Delta t + \tau $,
suppose $\Delta t\le t\le \Delta t + \tau $.
By the Heisenberg equation of motion, the position operator $Z(t)$ 
satisfies
\begin{equation}
 \frac{d}{dt} Z(t)
=\frac{1}{i \hbar }[Z(t),\frac{1}{2m}P(t)^2]=
\frac{1}{m} P(t).
\end{equation}

Thus, we have
\begin{equation}
Z(t) = Z(\Delta t ) + \frac{1}{m}\int_{\Delta t } ^{t} P(t') d t'.
\end{equation}
In contrast, $P(t)$ does not change since $[P(t),H(t)]=0$.
Consequently, we have
\begin{eqnarray}
Z(t) &=& Z(\Delta t ) + \frac{t-\Delta t}{m} P(\Delta t), \label{Z@@}\\
P(t) &=& P(\Delta t). \label{P@@}
\end{eqnarray}
Since $\sigma _z (t)$ and $\sigma _x (t)$ commute with $H(t)$, we have
\begin{equation}
\sigma _z(t) = \sigma _z(\Delta t), \quad \sigma _x(t) = \sigma _x(\Delta t).   \label{sigma_z@@}
\end{equation}

To describe the observables at time $t = \Delta t$ in terms of the observables at time $t = 0$,
suppose that $0\le t\le \De t$.
With the Heisenberg equations of motion,
we obtain
\begin{align}
\frac{d}{dt}Z(t)
&=\frac{1}{i\hbar}[Z(t),H(t)]=\frac{1}{m}P(t) \label{Heqz}
\end{align}
and
\begin{equation}
Z(\Delta t) = Z(0)+ \frac{1}{m}\int_{0}^{\Delta t}P(t) dt. \label{Z@}
\end{equation}
On the other hand, we have
\begin{equation}
\frac{d}{dt}P(t)= \frac{1}{i\hbar}[P(t),H(t)]
=-\mu B_1 \sigma _z (t).
\label{difP}
\end{equation}
Now $\sigma _z(t)$ commutes with Hamiltonian $H(t)$. Hence, we have 
\begin{equation}
\sigma _z (t) = \sigma _z (0). \label{sigma_z@}
\end{equation}
Consequently, we have 
\begin{eqnarray}
P(t) &=& P(0) - \mu B_1 t \sigma _z (0), \label{P@} \\
Z(t) &=&Z(0)+\frac{t}{m} P(0)-\frac{\mu B_1t^2}{2m}\sigma _z (0).
\end{eqnarray}
Therefore, we have
\begin{align}
Z(\Delta t + \tau )&=
Z(0)
+
\frac{\Delta t + \tau}{m}P(0)\notag\\
&\quad  -
\frac{\mu B_1 \Delta t}{m}
\left(
 \tau + \frac{\Delta t }{2}
\right)
\sigma _z(0), \label{ZjustaftermeasA}\\
P(\Delta t + \tau )
&=
P(0) - \mu B_1 \Delta t \sigma _z (0), \\
\sigma _z (\Delta t + \tau )
&=
\sigma _z (0). 
\end{align}

Next we calculate the $x$ and $y$ components of the spin of the particle at time $t = \Delta t + \tau $. 
Since the Hamiltonian $H(t)$ from time $t=\Delta t$ to time $\Delta t + \tau$ 
commutes with 
$\sigma _x(t)$ and $\sigma _y(t)$, we have
\begin{align}
\si_x(t)&=\si_x(\De t),   \label{eq:s-x}\\
\si_y(t)&=\si_y(\De t)   \label{eq:s-y}
\end{align}
if $\De t\le t\le\De t+ \ta$,
and it suffices to calculate $\sigma _x(\Delta t)$ and $\sigma _y(\Delta t)$.

Suppose $0\le t\le \De t$.
By the Heisenberg equations of motion we have
\begin{align}
\frac{d}{dt} \sigma _x(t) 
&=  \frac{1}{i\hbar } \left[ \sigma _x(t), H(t) \right] \notag \\
&=
\frac{1}{i\hbar }
\left[
 \sigma _x(t),
 \frac{P(t)^2}{2m}
 +
 \mu
  \left[
   B_0 + B_1 Z(t)
 \right]
 \sigma _z(t)
\right] \notag \\
&=
\frac{\mu}{i\hbar }
\left[
 B_0 + B_1 Z(t)
\right]
\left[
 -2 i \sigma _y(t) 
\right]
 \notag \\
&=
-\frac{2 \mu }{\hbar }
\left[
 B_0 + B_1 Z(t)
\right]
 \sigma _y(t).
\label{eq:HEM1}
\end{align}
Similarly, we have
\begin{align}
\frac{d}{dt} \sigma _y(t) 
&=\frac{1}{i\hbar } \left[ \sigma _y(t), H(t) \right] \notag \\
&=
\frac{1}{i\hbar }
\left[
 \sigma _y(t),
 \frac{P(t)^2}{2m}
 +
 \mu
 \left[
   B_0 + B_1 Z(t)
 \right]
 \sigma _z(t)
\right] \notag \\
&=
\frac{\mu}{i\hbar }
\left[
 B_0 + B_1 Z(t)
\right]
\left[
 2 i \sigma _x(t) 
\right]
 \notag \\
&=
\frac{2 \mu }{\hbar }
\left[
 B_0 + B_1 Z(t)
\right]
\sigma _x(t). 
\label{eq:HEM2}
\end{align}
Now let us introduce 
$\sigma_+$ and $\sigma _-$
by
\begin{eqnarray}
\sigma _+(t)
= \frac{1}{\sqrt{2}}
\left[
 \sigma _x(t) + i\sigma _y(t)
\right],\\
\sigma _-(t)
= \frac{1}{\sqrt{2}}
\left[
 \sigma _x(t) - i\sigma _y(t)
\right].
\end{eqnarray}
From Eqs. \eq{HEM1} and \eq{HEM2}, we have
\begin{equation}
\frac{d}{dt} \sigma _{\pm }(t)
=
\pm
\frac{2 \mu i}{\hbar}
\left[
 B_0 +
 B_1
 \left( U^\dag (t)Z(0)U(t) \right) 
\right]
\sigma _{\pm }(t). \label{Heisenbergeqofspinxywithkinetic}
\end{equation}
Let
\begin{equation}
\gamma _{\pm }(t)
=
U(t)
\sigma _{\pm} (t)
=
\exp
\left[
 \frac{H(0)}{i \hbar } t
\right] 
\sigma _{\pm} (t).
\end{equation}
The left-hand side (LHS) and right-hand side (RHS) 
of Eq.~(\ref{Heisenbergeqofspinxywithkinetic}) satisfy
\begin{align}
\mathrm{LHS}
&=
\frac{d}{dt} U(-t) \gamma _{\pm } (t) \notag \\
&= -\frac{H(0)}{i\hbar }U(-t) \gamma _{\pm }(t) + U(-t) \frac{d}{dt} \gamma _{\pm}(t),
\end{align}
\begin{align}
\mathrm{RHS}
&=
\pm
\frac{2 \mu i}{\hbar}
U^\dag(t)
\left[
 B_0 +
 B_1 Z(0) 
\right]
U(t)
U^\dag(t)
\gamma _{\pm }(t) \notag \\
&=
\pm
\frac{2 \mu i}{\hbar}
U(-t)
\left[
 B_0 +
 B_1 Z(0) 
\right]
\gamma _{\pm }(t).
\end{align}
Hence, we have
\begin{equation}
\frac{d}{dt} \gamma _{\pm }(t)
=
\left(
 \frac{H(0)}{i\hbar } \pm \frac{2 \mu i}{\hbar }
 \left[
  B_0 + B_1Z(0)
 \right]
\right)
\gamma _{\pm }(t).
\end{equation}
The solution of the above differential equation is
given by
\begin{equation}
\gamma _{\pm }(t)
=
\exp
\left(
 \frac{it}{\hbar }
 \left\{
  -H(0) \pm 2 \mu
  \left[
   B_0 + B_1 Z(0)
  \right]
 \right\}
\right)
\gamma _{\pm }(0).
\end{equation}
Since $\gamma _{\pm }(0) = \sigma _{\pm}(0)$, we have 
\begin{multline}
\sigma _{\pm }(t)
=
\exp
\left(
 \frac{it}{\hbar } H(0)
\right) \\
\times
\exp
\left(
 \frac{it}{\hbar }
 \left\{
  -H(0) \pm 2 \mu
  \left[
   B_0 + B_1 Z(0)
  \right]
 \right\}
\right)
\sigma _{\pm } (0). \label{timeevolutionofsigmapm}
\end{multline}
Using the Baker-Campbell-Hausdorff formula
\cite{Bak05}
we have
\begin{multline}
\exp (A) \exp (B)
=
\exp
\Bigl\{
 (A+B)
 +
 \frac{1}{2} \left[ A ,B \right] \\
 +
 \frac{1}{12}
 \left(
  \left[ \left[ A ,B \right], B\right]
  -
  \left[ \left[ A ,B \right], A\right]
 \right)
 +
 \cdots
\Bigr\}{.}
\label{eq:BCH}
\end{multline}
Hence, for
\begin{eqnarray}
A
&=&
\frac{it}{\hbar} H(0), \\
B
&=&
\frac{it}{\hbar}
\left\{
 -H(0) \pm 2 \mu
 \left[
  B_0 + B_1 Z(0)
 \right]
\right\},
\end{eqnarray}
we have
\begin{align}
[A, B]
&=
\left[
 \frac{it}{\hbar } H(0),
 \frac{it}{\hbar }
 \left\{
  -H(0) \pm 2 \mu \left[ B_0 + B_1 Z(0) \right]
 \right\}
\right] \notag \\
&=
-\frac{t^2}{\hbar ^2}
\left[
 \frac{1}{2m}P(0)^2,
 \pm 2 \mu 
  \left[ B_0 + B_1 Z(0) \right]%
\right] \notag \\
&=
\pm
\frac{2i \mu B_1 t^2}{m \hbar }
P(0),
\end{align}
\begin{align}
\left[[A, B], A \right]
&=
\left[
 \pm \frac{2i\mu B_1 t^2}{m \hbar } P(0),
 \frac{it}{\hbar } H(0)
\right] \notag \\
&=
\mp \frac{2 \mu B_1 t^3}{m \hbar ^2}
\left[
 P(0),
 \mu 
  \left[B_0 + B_1 Z(0) \right]%
 \sigma _z(0) 
 \right] \notag \\
&=
\pm \frac{2i \mu ^2 B_1^2 t^3}{m \hbar } \sigma _z(0),
\end{align}
\begin{align}
 &\left[[A, B], B \right] \notag \\
&=
\left[
 \pm \frac{2i\mu B_1 t^2}{m \hbar } P(0),
 \frac{it}{\hbar }
 \left\{
 -H(0) \pm 2 \mu \left[ B_0 + B_1 Z(0) \right]
 \right\}
\right] 
\notag \\
&=
\mp \frac{2i \mu ^2 B_1^2 t^3}{m \hbar } \sigma _z(0) \notag \\
& \quad \mp \frac{2 \mu B_1 t^3}{m \hbar ^2}
\left[
 P(0),
 \pm 2 \mu
  \left[B_0 + B_1 Z(0) \right]
  \sigma _z(0)
\right] \notag\\
&=
\frac{2i \mu ^2 B_1^2 t^3}{m \hbar } 
 \left[ 2 \mp \sigma _z(0) \right].
\end{align}
The commutators of the higher orders, denoted by an ellipsis 
in \Eq{BCH}, are $0$ since the third commutators $[[A,B], A]$ and $[[A,B], B]$ 
commute with $A$ and $B$, respectively.

Let
\begin{eqnarray}
R(t)&=&
\frac{\mu ^2 B_1^2 t^3 }{3m\hbar }, \\
S(t)&=&
\frac{2 \mu t }{\hbar }
\left[
 B_0 + B_1 \left( Z + \frac{t}{2m}P \right)
\right].
\end{eqnarray}
We have
\begin{equation}
\sigma _{\pm } (t)
=
\exp
i
\left\{
[ R(t) \pm S(t)] \1 \mp R(t)\sigma _z(0)
\right\}
\sigma _{\pm } (0). 
\end{equation}
Since
\begin{eqnarray}
\sigma _+ (0)
&=&
\frac{1}{\sqrt{2}}
\left[
 \sigma _z (0) + i \sigma _y (0)
\right]
=
\left(
 \begin{array}{cc}
  0 & \sqrt{2} \\
  0 & 0
 \end{array} 
\right),\\
\sigma _- (0)
&=&
\frac{1}{\sqrt{2}}
\left[
 \sigma _z (0) - i \sigma _y (0)
\right]
=
\left(
 \begin{array}{cc}
  0        & 0 \\
  \sqrt{2} & 0
 \end{array}
\right),
\end{eqnarray}
we have
\begin{align}
&\sigma _+ (t) \notag \\
&=
\left(
 \begin{array}{cc}
  \exp [iS(t)]         & 0             \\
  0                & \exp i[S(t) + 2R(t) ]
 \end{array}
\right)
\left(
 \begin{array}{cc}
  0 & \sqrt{2} \\
  0 & 0
 \end{array}
\right) \notag \\
&=
\exp [i S(t)]  \sigma _+ (0),
\end{align}
\begin{align}
 &\sigma _- (t) \notag \\
&=
\left(
 \begin{array}{cc}
  \exp  \left\{ i[-S(t) + 2R(t)] \right\} & 0                 \\
                                         0               & \exp [-iS(t)]
 \end{array}
\right)
\left(
 \begin{array}{cc}
  0        & 0 \\
  \sqrt{2} & 0
 \end{array}
\right) \notag \\
&=
\exp  [-iS(t)]  \sigma _- (0). 
\end{align}
Therefore, $\sigma _x (t)$ and $\sigma _y (t)$ from time $t=0$ to time $t=\Delta t$ are
\begin{align}
\sigma _x (t)
&=
\frac{1}{\sqrt{2}}
\left[
 \sigma _+ (t) + \sigma _- (t)
\right] 
\notag\\
&=
\left(
 \begin{array}{cc}
  0        & \exp [iS(t)] \\
  \exp [-iS(t)] & 0
 \end{array}
\right), \label{sigma_zjustaftermeasA}
\end{align}
\begin{align}
\sigma _y (t)
&=
-\frac{i}{\sqrt{2}}
\left[
 \sigma _+ (t) - \sigma _- (t)
\right]
\notag\\
&=
\left(
 \begin{array}{cc}
  0          & -i \exp [iS(t)] \\
  i \exp [-iS(t)] & 0
 \end{array}
\right).
\end{align}

\section{Supremum of the function $W _{\lambda }(t)$}
\label{se:App_Sup}
Let us consider the supremum of the function in 
Sec.~\ref{squeezeded}, 
\begin{equation}
W_{\xi _{\lambda }}(\tau )
=
\alpha
\left(
 \tau \!+\! \frac{\Delta t }{2}
\right)
\left[
 a\!+\!b(\Delta t \!+\! \tau )\!+\!c(\Delta t \!+\! \tau )^2
\right]^{-1/2} \label{W}.
\end{equation}
Here we 
set 
$\alpha = \dfrac{\mu B_1 \Delta t}{\sqrt{2}m}$,
$a = \left\langle Z^2 \right\rangle$,
$b = \dfrac{\left\langle \left\{ Z , P \right\} \right\rangle }{m}$, and
$c = \dfrac{\left\langle P ^2\right\rangle }{m^2}$.
The derivative of function $W_{\xi_{\lambda }}(\tau )$ is
\begin{align}
&\frac{d}{d \tau }W _{\lambda}(\tau )  \notag \\
&=
\frac{\alpha }{4}
\left[
 a+b(\Delta t + \tau )+c(\Delta t + \tau )^2
\right]^{-3/2} \notag \\
&\times
\left[
 2(b + c\Delta t)(\Delta t + \tau ) + 4a + b\Delta t
\right].
\end{align}
Hence, $W_{\xi _{\lambda }} (t)$ assumes the maximum value at 
$\tau =\tau_0 =-\dfrac{4a+3b\Delta t+2c\Delta t ^2}{2(b+c\Delta t)} \geq 0$ if
the following conditions hold:
(i)  $W'(0) >0 $ and
(ii) $2b + 2c\Delta t <0$.

Condition (i) holds automatically. In fact, (i) is equivalent to the condition
\begin{equation}
4a + 3b\Delta t + 2c\Delta t^2 \geq 0.
\end{equation}
Now let us consider the function
\begin{equation}
f(t)=4a + 3b t + 2c t^2.
\end{equation}
This function assumes the minimum value at $t= -\frac{3b}{4c}$,
\begin{align}
f(t)& \geq 
f\left(-\dfrac{3b}{4c}\right) \notag \\
&=
\frac{32ac-9b^2}{8c} \notag \\
&=
\frac{9}{8c}(4ac - b^2)-\frac{4ac}{8c} \notag \\
&\geq
\frac{9 \hbar^2}{8cm^2}-\frac{\hbar ^2}{8cm^2} \notag \\
&=
\frac{\hbar^2}{cm^2} \notag \\
&> 0.
\end{align}
Therefore, condition (i) is satisfied automatically.
Here we use
the Schr\"{o}dinger inequality (\ref{schr_ineq}). 
Hence, if condition (ii) holds, 
the function $W _{\lambda }(\tau )$ assumes the 
maximum value at $\tau = \tau _0 \geq 0$.
The maximum value of $W_{\xi _{\lambda }}(\tau )$ for $\tau \geq 0$ is
\begin{align}
W_{\xi _{\lambda }}(\tau _0)
&=
-\alpha\frac{4a + 2b \Delta t + c \Delta t^2}{2(b + c \Delta t)} \notag \\
&\times
\left[
 a + b(\Delta t + \tau _0) + c (\Delta t + \tau _0)^2
\right]^{-1/2} \notag \\
&=
\alpha\left(4a + 2b \Delta t + c \Delta t^2\right)^{1/2}
(4ac-b^2)^{-1/2} \notag \\
&=
\frac{2\alpha m}{\hbar }
\left[ a + b \frac{\Delta t}{2}  + c \left(\frac{\Delta t}{2}\right)^2 \right]^{1/2} \notag \\
&=
\frac{\sqrt{2}\mu B_1 \Delta t}{\hbar}
\left\langle
 \left(
  Z + \frac{\Delta t}{2m} P
 \right)^2
\right\rangle _{\xi _\lambda }^{1/2}. \label{Wmaxmum}
\end{align}
If condition (ii) does not hold, 
the function $W_{\xi _{\lambda }}(\tau )$ increases monotonically 
and we have
\begin{equation}
\sup _{\tau \geq 0 } 
W_{\xi _{\lambda }}(\tau ) 
= 
\lim_{\tau \to \infty }
W_{\xi _{\lambda }}(\tau )
=
\frac{\mu B_1 \Delta t }{\sqrt{2\langle P^2\rangle_{\xi _\lambda } }}.
\end{equation}

\end{document}